\documentclass[sigconf]{acmart}
\acmConference[CAIN 2024]{3rd International Conference on AI Engineering — Software Engineering for AI}{April 2024}{Lisbon, Portugal}

\copyrightyear{2024}
\acmYear{2024}
\setcopyright{acmlicensed}
\acmConference[CAIN 2024]{Conference on AI Engineering - Software Engineering for AI}{April 14--15, 2024}{Lisbon, Portugal}
\acmBooktitle{Conference on AI Engineering - Software Engineering for AI (CAIN 2024), April 14--15, 2024, Lisbon, Portugal}
\acmDOI{10.1145/3644815.3644989}
\acmISBN{979-8-4007-0591-5/24/04}

\usepackage{subcaption}
\usepackage{graphicx}

\usepackage[font=small,labelfont=bf]{caption}
\usepackage{subcaption}

\usepackage{hyperref}

\usepackage{lineno}

\usepackage{amsthm}
\usepackage{mdframed}
\usepackage{centernot}
\usepackage{comment}

\usepackage{wrapfig,lipsum,booktabs}

\newcommand{\Hh}{\mathcal{H}}

\newcommand{\Xx}{\mathcal{X}}
\newcommand{\Yy}{\mathcal{Y}}

\newcommand{\Dd}{\mathcal{D}}

\newcommand\vd[2]{d_{i, p}}
\newcommand{\set}[1]{\left\{ #1 \right\}}

\usepackage{tikz}

\newtheorem{definition}{Definition}[section]

\definecolor{gold}{rgb}{0.99,0.78,0.07}
\usetikzlibrary{arrows,shapes,snakes,automata,backgrounds,positioning,decorations.pathmorphing,
	decorations.markings,calc}

\tikzstyle{dtreenode}=[draw=blue!10!gray,rounded rectangle, minimum size=5mm,fill=blue!10!white]
\tikzstyle{dtreeleaf}=[draw=black!60,minimum width=1cm,minimum height=0.4cm,rectangle,fill=blue!50!white]
\tikzset{every loop/.style={looseness=7}}
\tikzset{
	gluon/.style={decorate,draw=black,
		decoration={coil,amplitude=1pt, segment length=5pt}}
}
\tikzset{
	gluon1/.style={decorate,draw=black,
		decoration={coil,amplitude=3pt, segment length=3pt}}
}
\tikzset{
	gluonew/.style={decorate,draw=black,
		decoration={coil,amplitude=1pt, segment length=2pt}}
}

\tikzset{bicolor/.style args={#1 and #2 and #3}{
		path picture={
			\tikzset{rounded corners=0}
			\fill [#1] (path picture bounding box.south west)
			rectangle
			($(path picture  bounding box.north west)!#3!(path picture bounding
			box.north east)$);
			\fill [#2]
			($(path picture bounding box.south west)!#3!(path picture bounding
			box.south east)$)
			rectangle (path picture bounding box.north east);
}}}

\tikzset{tricolor/.style args={#1 and #2 and #3 and #4 and #5}{
		path picture={
			\tikzset{rounded corners=0}
			\fill [#1] (path picture bounding box.south west)
			rectangle
			($(path picture  bounding box.north west)!#4!(path picture bounding
			box.north east)$);
			\fill [#2]
			($(path picture bounding box.south west)!#4!(path picture bounding
			box.south east)$)
			rectangle
			($(path picture  bounding box.north west)!#5!(path picture bounding
			box.north east)$);
			\fill [#3]
			($(path picture bounding box.south west)!#5!(path picture bounding
			box.south east)$)
			rectangle (path picture bounding box.north east);
}}}

\usepackage[many]{tcolorbox}
\tcbuselibrary{listings,skins}

\lstdefinestyle{mystyle}{
  xleftmargin=0pt,
   basicstyle={\footnotesize\ttfamily},
   aboveskip=3mm,
   belowskip=3mm,
   keywordstyle=\bfseries,
   showstringspaces=false,
  escapechar=?,
  language=Java
}
\definecolor{code_indent}{HTML}{CCCCCC}

\newtcblisting{mylisting}[2][]{
  arc=0pt, outer arc=0pt,
  listing only,
  listing style=mystyle,
  title={\large #2},
  #1
}

 \definecolor{dkgreen}{rgb}{0,0.6,0}
 \definecolor{gray}{rgb}{0.5,0.5,0.5}
 \definecolor{mauve}{rgb}{0.58,0,0.82}

\definecolor{cadmiumgreen}{rgb}{0.0, 0.42, 0.24}
\definecolor{verde}{rgb}{0.25,0.5,0.35}
\definecolor{jpurple}{rgb}{0.5,0,0.35}
\definecolor{darkgreen}{rgb}{0.0, 0.2, 0.13}

\usepackage[ruled,vlined,linesnumbered,lined]{algorithm2e}
\usepackage{algpseudocode}

\usepackage{mathtools,xparse}

\usepackage{tabu}
\usepackage{multirow}

\usepackage{pifont}
\usepackage{enumerate}
\usepackage[shortlabels]{enumitem}

\usepackage{pgfplotstable}
\usepackage{pgfplots}

\usepackage[noframe]{showframe}
\usepackage{framed}

 {\endMakeFramed}
 \definecolor{shadecolor}{gray}{0.85}

\definecolor{bgblue}{RGB}{245,243,253}
\definecolor{ttblue}{RGB}{91,194,224}
\definecolor{mygray}{gray}{0.6}
\definecolor{aliceblue}{rgb}{0.94, 0.97, 1.0}

\mdfdefinestyle{mystyle}{%
  rightline=true,
  innerleftmargin=10,
  innerrightmargin=10,
  outerlinewidth=3pt,
  topline=false,
  rightline=false,
  bottomline=false,
  skipabove=\topsep,
  skipbelow=false%\topsep
}

\newtcolorbox{myboxi}[1][]{
  breakable,
  title=#1,
  colback=white,
  colbacktitle=white,
  coltitle=black,
  fonttitle=\bfseries,
  bottomrule=0pt,
  toprule=0pt,
  leftrule=3pt,
  rightrule=3pt,
  titlerule=0pt,
  arc=0pt,
  outer arc=0pt,
  colframe=black!50,
}

\newtcolorbox{myboxii}[1][style=mystyle]{
  breakable,
  freelance,
  colback=white,
  colbacktitle=white,
  coltitle=black,
  fonttitle=\bfseries,
  bottomrule=0pt,
  boxrule=0pt,
  colframe=white,
  after skip=0pt,
  overlay unbroken and first={
    \draw[white!75!black,line width=3pt]
    ([yshift=-9pt]frame.north west) --
    ([yshift=9pt]frame.south west);
  },
  }
\usepackage{colortbl}

\begin{document}

\title[WCCT of ML algorithms via EVT]{Worst-Case Convergence Time of ML Algorithms via \\ Extreme Value Theory}

\author{Saeid Tizpaz-Niari}
\affiliation{%
  \institution{University of Texas at El Paso}
  % \streetaddress{1 Th{\o}rv{\"a}ld Circle}
  \city{El Paso, TX}
  \country{USA}}
\email{saeid@utep.edu}

\author{Sriram Sankaranarayanan}
\affiliation{%
  \institution{University of Colorado Boulder}
  % \streetaddress{1 Th{\o}rv{\"a}ld Circle}
  \city{Boulder, CO}
  \country{USA}}
\email{srirams@colorado.edu}

\begin{abstract}
This paper leverages the statistics of extreme values to predict the worst-case
convergence times of machine learning algorithms. 
Timing is a critical non-functional property of ML systems, and providing the worst-case
converge times is essential to guarantee the availability of ML and its services.
However, timing properties such as worst-case convergence times (WCCT) are difficult to verify since
(1) they are not encoded in the syntax or semantics of underlying programming languages of AI,
(2) their evaluations depend on both algorithmic implementations and underlying systems, and
(3) their measurements involve uncertainty and noise. 
Therefore, prevalent formal methods and statistical models fail to provide
rich information on the amounts and likelihood of WCCT.

Our key observation is that the timing information we seek represents the extreme tail of execution times.
Therefore, extreme value theory (EVT), a statistical discipline that focuses on understanding and predicting the distribution of extreme values in the tail of outcomes, provides an ideal framework to model and analyze WCCT in the training
and inference phases of ML paradigm. Building upon the mathematical tools from EVT,
we propose a practical framework to predict the worst-case timing properties of ML.
Over a set of linear ML training algorithms,
we show that EVT achieves a better accuracy for predicting WCCTs than relevant statistical methods
such as the Bayesian factor. On the set of larger machine learning training algorithms and deep neural network inference,
we show the feasibility and usefulness of EVT models to accurately predict WCCTs, their expected return
periods, and their likelihood.

\end{abstract}

\maketitle

\section{Introduction}
\label{sec:intro}
Machine learning (ML) has been significantly integrated into modern software developments where they are routinely leveraged to assist in safety-critical decision-making such as autonomous cars~\cite{levinson2011towards}, medical diagnosis~\cite{ciresan2012deep}, malware detection~\cite{yuan2014droid}, information leak~\cite{tizpaz2019efficient}, and aircraft collision avoidance systems~\cite{julian2016policy}. Recently, neuron-based foundational AI models such as ChatGPT showed great performance on some of the most challenging programming tasks such as synthesizing programs from a high-level natural language specification and even outperformed domain experts (e.g., lawyer and physician) in answering questions. 

Such wide adoption of ML techniques comes with concerns about reliability, accountability, privacy, fairness, greenness, etc. One substantial concern, especially due to the rapid adaptation of large language models, is the environmental risks~\cite{schwartz2020green} of training ML models where the computations required for deep neural networks (DNNs) have been doubling every few months~\cite{AI-computation}. The software engineering community considers the timing analysis as a critical 
non-functional property and has spent significant efforts to verify, validate, and analyze the timing properties of software.
However, there are no systematic methods to predict and provide guarantees on the worst-case computation times of ML-based software systems to reduce their environmental risks, improve their runtime performance, and increase their availability.
Similar to traditional software, one challenge is that timing is not encoded as a part of the syntax or semantics of underlying programming languages. Furthermore, it is a product of both software and platforms that execute the software. Modern data-driven software brings new challenges to the analysis since the computation times significantly depend on the characteristics of training data, in addition to the architecture of ML models, their hyperparameters, GPUs, etc.  

Due to these challenges in the static verification of timing properties, one approach is to explore dynamic analysis techniques and provide statistical guarantees on the estimation of worst-case convergence times (WCCT), i.e., time taken to reach a desirable state such as a loss value below a threshold during training. Statistical model checking via hypothesis testing is a common method to provide such statistical guarantees. In doing so, one might come up with two hypotheses where the null hypothesis is a predicate that the execution times are below a threshold and the alternative hypothesis is the negation of such predicate. Then, they can use a Bayesian factor testing such as Jeffreys~\cite{10.1007/978-3-642-03845-7_15} to accept the null hypothesis with very strong evidence if the experiments witness $K \geq 90$ sequential \texttt{true} evaluations of the null hypothesis. Similarly, the rule of three~\cite{jovanovic1997look} can provide 95\% confidence intervals [0, 3/K] on the likelihood of observing a \texttt{false} evaluation given that we have observed $K\geq30$ \texttt{true} evaluations. However, such hypothesis testing provides limited information and may fail to provide richer quantitative information about the severity (amounts), return periods, and likelihood of worst-case convergence times.

In this paper, we propose a method and a tool to provide a quantitative estimation of worst-case convergence times based on extreme value theory (EVT)~\cite{de2006extreme}. While statistical theories such as the central limit theorem focus on the expected quantities of random variables, they often overlook unusually rare quantities. The EVT overcomes this problem by focusing on the extremely rare events and high (low) quantities in the random observations. Prior works have significantly leveraged EVT to bound the worst-case execution times of programs in the embedded and real-time systems where the motivations are the uncertainty of underlying hardware systems~\cite{lu2011new,cucu2012measurement,hansen2009statistical}. In addition, EVT has been used to find rare bugs in circuit design~\cite{singhee2007statistical,8342220}. To the best of our knowledge, this is the first work to study the feasibility, scalability, and usefulness of EVT for providing probabilistic bounds on the convergence times of ML algorithms.

\begin{quote}
\emph{Our key observation is that the worst-case convergence times (WCCT) of data-driven software over independent random samples represent the maximum over a set of random variables. Thus, if valid, EVT can provide useful information to model and predict the WCCT of ML algorithms with probabilistic guarantees.}    
\end{quote}

Our experiments include both micro-benchmarks as well as realistic machine learning algorithms. Over a set of linear training benchmarks, we found that it is feasible to model the worst-case computation times via EVT, and it significantly improves the accuracy of WCCT predictions as compared to the
baseline Bayes factor~\cite{jha2009bayesian,sankaranarayanan2013static} in 83\% of cases.
Given at most 261 samples, the EVT method is able to predict the actual WCCTs of the next 10,000 queries
with more than 75\% accuracy (compared to the baseline) in 40\% of cases. 
 
Over $4$ popular ML training algorithms (logistic regression, decision trees, Gaussian process, and discriminant analysis),
we found that EVT is scalable and accurately predicts the WCCT of training in 57\% of cases. Over $3$ deep neural
network models as controllers for cyber-physical systems, we found that EVT can accurately predict the WCCT of inference
convergences in 75\% of cases. Our observations include: i) EVT might be a more useful tool in the inference stage compared
to the training stage; and ii) EVT extrapolations become more accurate in the longer horizon (e.g., it is more accurate
to predict the WCCT up to 10K queries as compared to 500 queries). 
In summary, we make the following contributions:
\begin{itemize}
    \item A \textit{feasibility} study of applying extreme value theory for reasoning about the worst-case convergence times of data-driven applications,
    \item A \textit{quantitative} statistical method that measures the severity, period, and likelihood of extremely rare computation times for the convergence, and
    \item A large set of \textit{experiments} that show the usefulness and scalability of our approach in adapting EVT for ML training processes and DNN-based inference. 
\end{itemize}

\section{Background}
\label{sec:prelim}
Extreme value theory~\cite{coles2001introduction} is a statistical branch that deals
with the analysis of extreme events in a random process.
Given a set of independent and identically distributed random variables $\set{z_1,\ldots,z_n}$, the extreme
value theory is concerned with the min/max statistics of a random process, i.e., $M_n = \max(\set{z_1,\ldots,z_n})$
or $M_n = \min(\set{z_1,\ldots,z_n})$ as $n \to \infty$. 

Under some mild assumptions, it has been proved (e.g., see Leadbetter et al.~\cite{leadbetter2012extremes}) that $Pr[(M_n-b_n)/a_n < z] \to G(z)$ as $n \to \infty$ and $G$ belongs to a family of distributions called the \emph{generalized extreme value (GEV)} family. Each such distribution has the CDF of
\[
G(z) = \exp\left\{-\left[1 + \xi\left(\frac{z - \mu}{\sigma}\right) \right]^{-1/\xi}\right\},
\]
defined over $\{z: 1 + \xi(z - \mu)/\sigma > 0\}$. 
The model has three parameters: a location parameter $-\infty<\mu<+\infty$, a scale parameter
$\sigma > 0$, and a shape parameter $-\infty < \xi < +\infty$. Type I, known as the Gumbel family, defines a subset of GEV distribution when $\xi \to 0$. The tail behavior of type I, $z_{+}$, has infinite support, but the density of GEV decays exponentially (guarantees are feasible up to a bounded). For type II, $\xi > 0$ and $z_{+}$ have infinite support, decaying polynomially (limited guarantees are feasible). For a special case where $\xi \geq 1$, the mean of GEV is infinite, decaying logarithmically, so no statistical guarantee is feasible. Finally, for type III, $\xi < 0$ and $z_{+}$ has finite support. In this case, the statistical guarantees on the worst-case outcomes are feasible for a long horizon.

\vspace{0.25em}
\noindent \textit{Generalized Pareto Distribution.} 
There are two basic approaches to infer the parameters of GEV distributions: block maximum and threshold approach. The block maximum approach divides samples into blocks of the same size and uses the maximum of each block as the extreme value. Since such an approach is more appropriate for seasonal data, in this paper, we use the threshold approach where extreme events that exceed some high threshold $u$, i.e., $\{x_i: x_i > u\}$, are extreme values. Labeling these exceedances by
$\{\delta_{(1)},\ldots,\delta_{(k)}\}$, we define threshold excesses by $\delta_j = x_j - u$ for $1 {\leq} j {\leq} k$. It follows that if $Pr[M_n < z] \to G(z)$, then for large enough $u$, the distribution function of ($T-u$), conditional on $T>u$, is approximately 
\[
H(t) = 1 - \left(1 + \frac{{\xi}t}{\hat{\sigma}} \right)^{-1/\xi}
\]
where $t>0$ and $\hat{\sigma} = \sigma + \xi(u - \mu)$~\cite{coles2001introduction}.
This distribution is known as \textit{generalized Pareto} distribution. The implication of shape parameter $\xi$ is the
same as $G(z)$, as a special case of GEV distribution.

\vspace{0.5em}
\noindent \textit{Threshold Selection.} 
A proper choice of threshold value $u$ is critical to analyze the
behavior of extreme value distributions. Low values of threshold $u$ might
include non-tail samples and lead to mixture distributions that violate
the asymptotic basis of the model. On the other hand, high values of
threshold $u$ might include only a few tail samples and lead to low
confidence in the model due to high variance. Therefore, it is critical
to be confident on the threshold value to provide any guarantees on the
worst-case fairness behaviors. 

\vspace{0.5em}
\noindent \textit{Return Levels.}
It is often convenient to model extreme value distributions
using return levels. The inverse of the probability density function of GEV at
probability $p$, is the \textit{return level} $\delta_p$,  associated with the
\textit{return period} $1/p$. Therefore, the level $\delta_p$ is expected
to be exceeded on average once every $1/p$ period of time.
A \textit{return level} is represented with
{$(m,\delta_m)$} where $m$ is the time period (e.g., the number of queries with the ML model) and the level $\delta_m$ is the expected extreme value during the $m$ period (e.g., expected worst-case execution times in the next $m$ queries).

\section{Overview}
\label{sec:overview}

\begin{figure*}[tbp!]
    \centering
    \begin{subfigure}[b]{0.24\textwidth}
        \includegraphics[width=\textwidth]{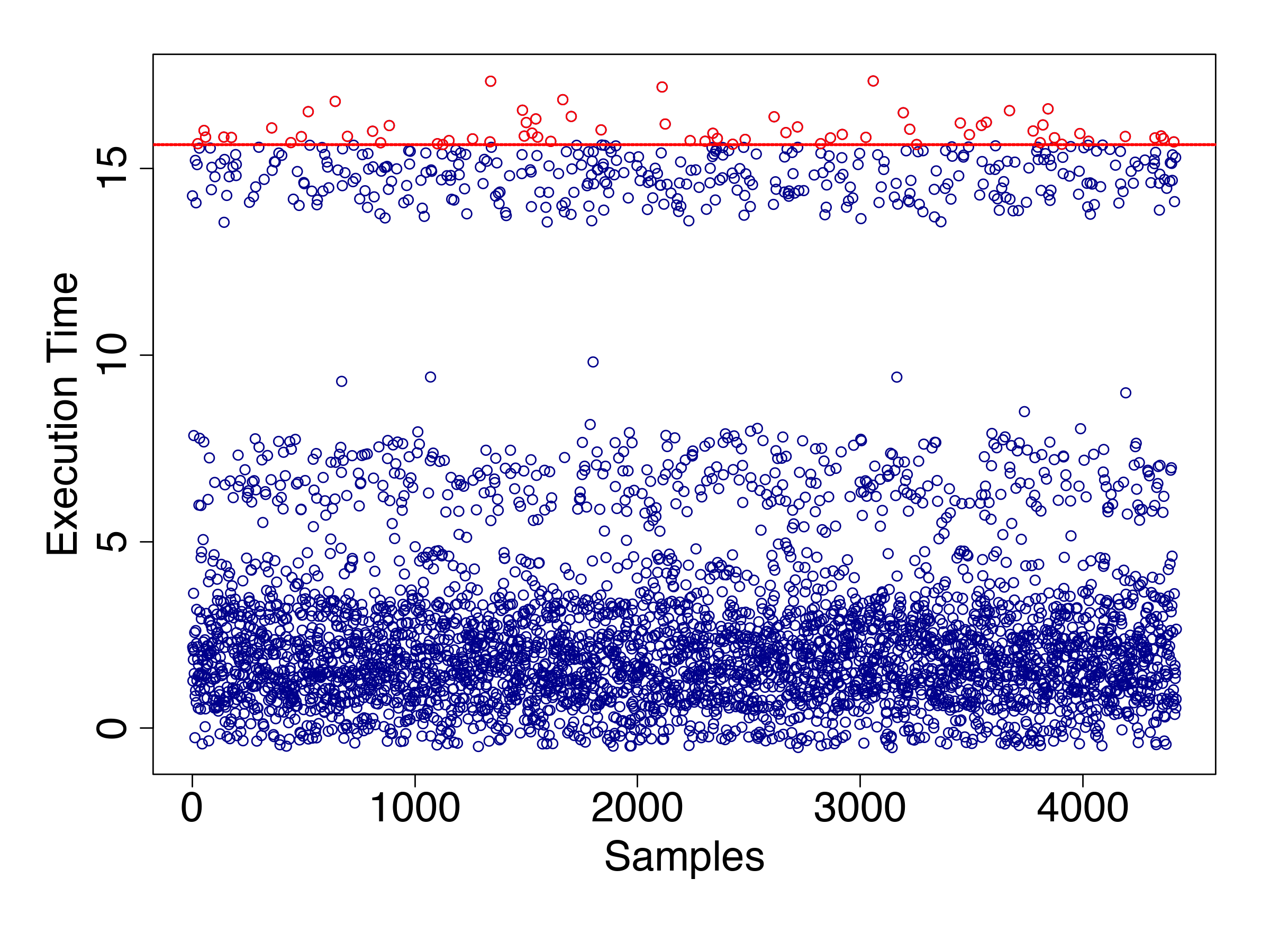}
        \caption{}
        \label{fig:SB_threshold}
    \end{subfigure}
    \begin{subfigure}[b]{0.24\textwidth}
        \includegraphics[width=\textwidth]{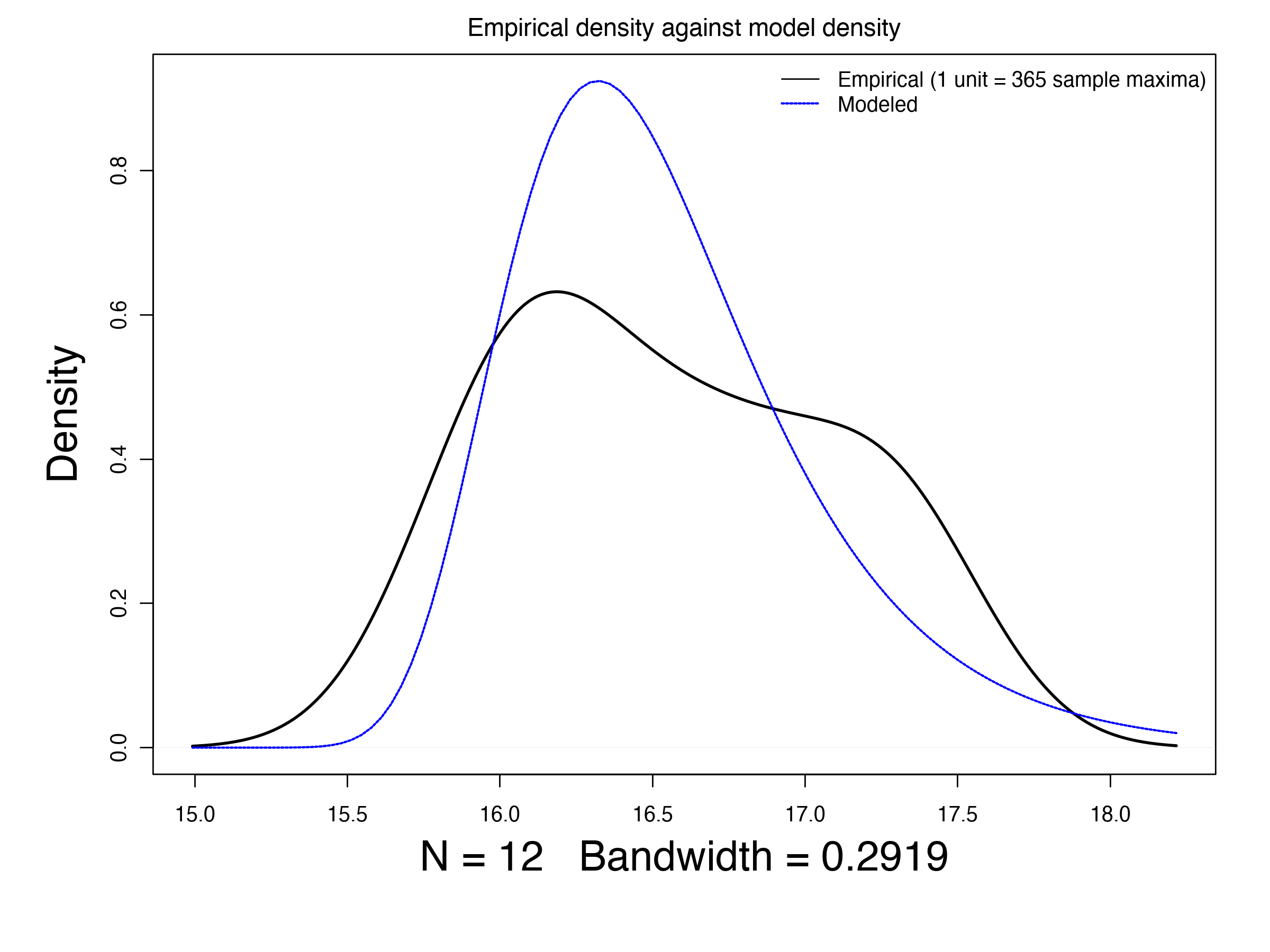}
        \caption{}
        \label{fig:SB_density}
    \end{subfigure}
    \hspace{0.5em}
    \begin{subfigure}[b]{0.24\textwidth}
        \includegraphics[width=\textwidth]{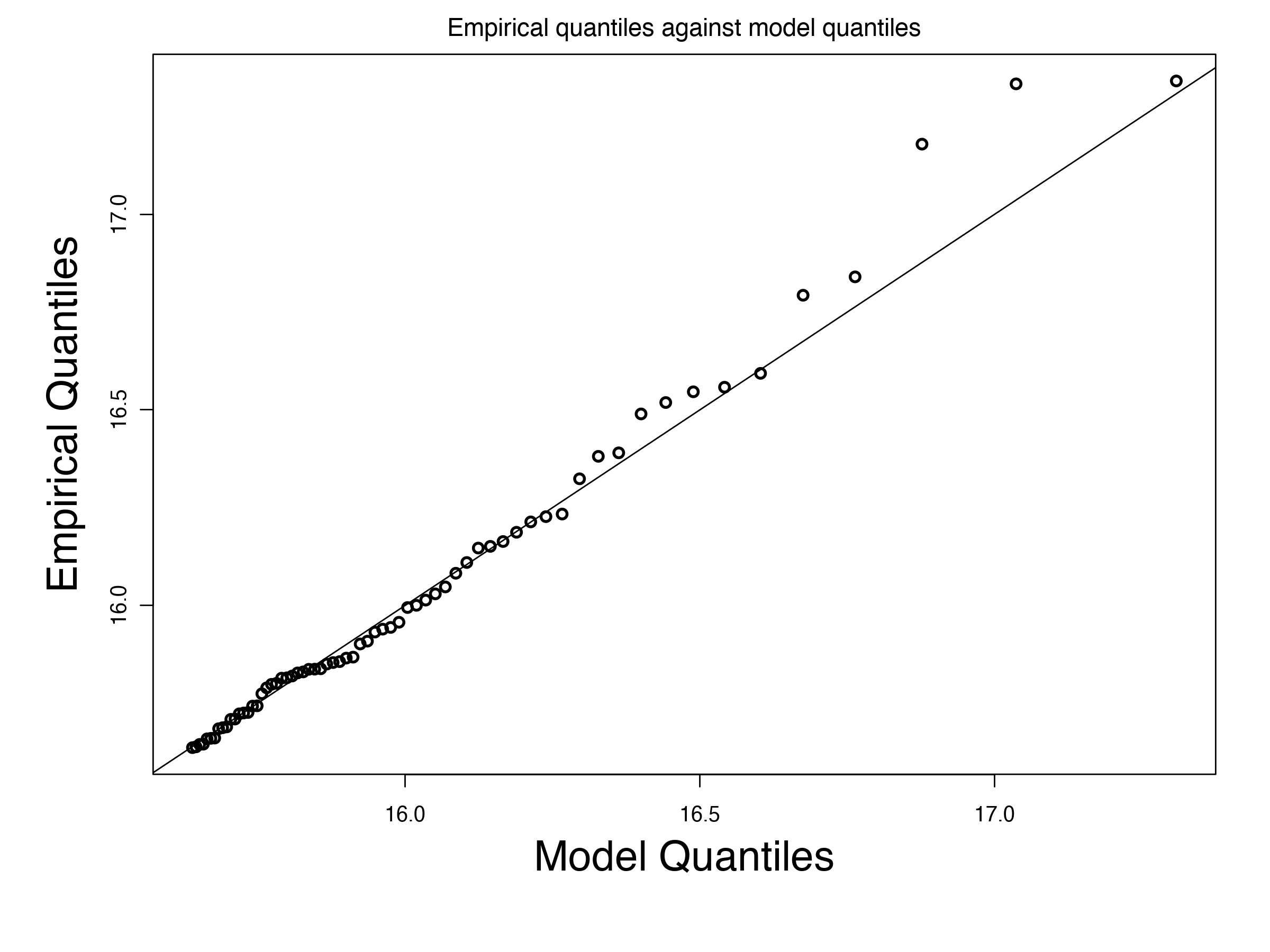}
        \caption{}
        \label{fig:SB_qq}
    \end{subfigure}
    \begin{subfigure}[b]{0.24\textwidth}
        \includegraphics[width=\textwidth]{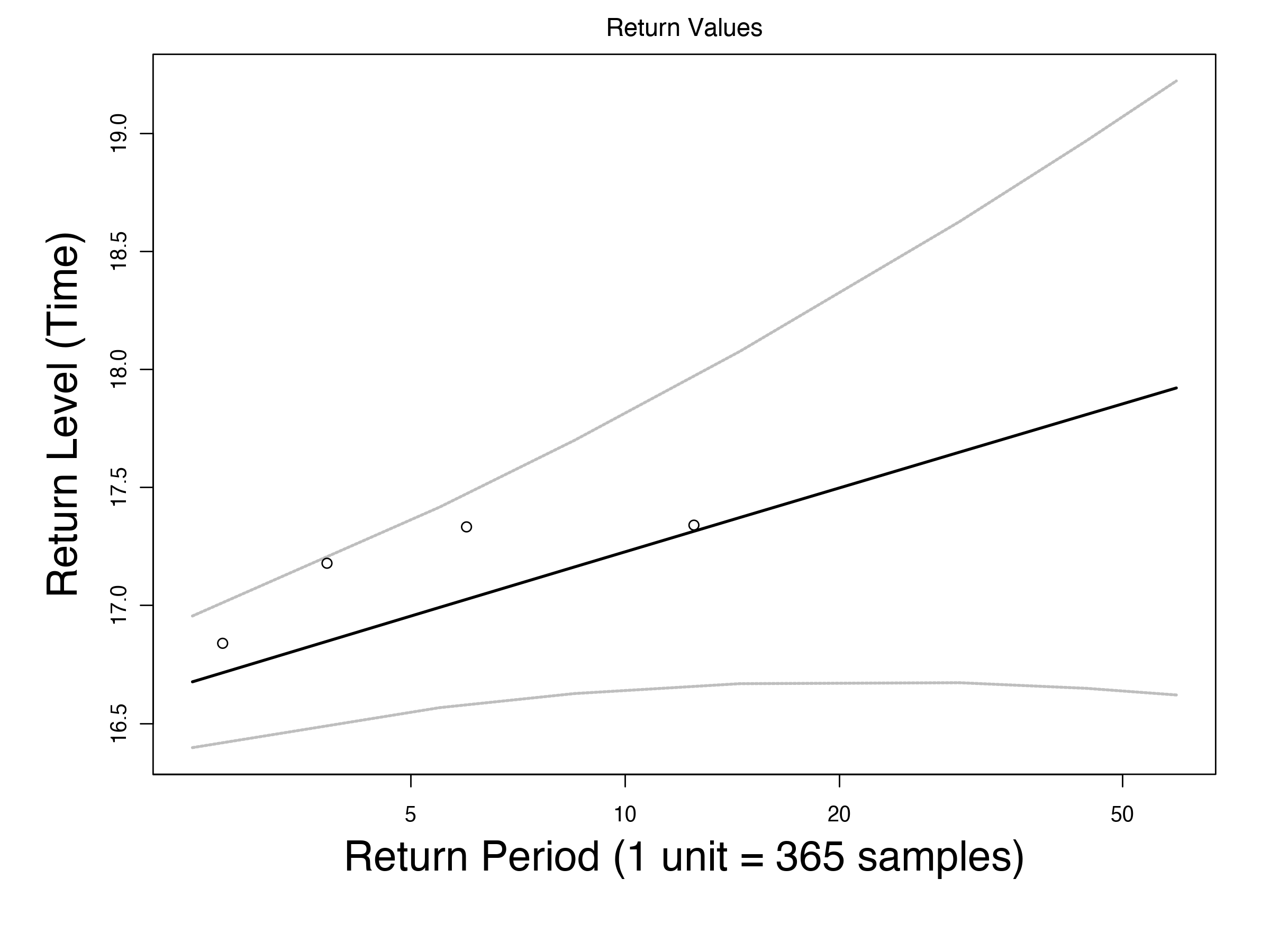}
        \caption{}
        \label{fig:SB_return_level}
    \end{subfigure}
    \caption{Overview Example. (a) The computation times of applying photo filters with a threshold of 15.6 seconds.
    (b) the density plot for GEV distribution (blue) vs. empirical model (black), (c) quantile Plot for the execution time of the photo algorithm,
    (d) m-return level plot for the computation times of the algorithm with expected values and their 95\% CI.
}
\end{figure*}

To illustrate the process of extreme value analysis and show the efficacy of EVT in analyzing the worst-case execution times, we use a simple example of a mock social network server~\cite{snapbuddy} that applies different actions on the profile of users. We send requests to apply different types of photo filters on the images that have different convergence times. 

\vspace{0.5 em}
\noindent \textbf{Test Cases.} 
We randomly generate inputs and collect the computation times of applying photo filters on the public image of users.
Overall, we collected 4,419 response times. 

\vspace{0.5 em}
\noindent \textbf{Characteristics of EVT distribution.} 
Figure~\ref{fig:SB_threshold} shows the computation times where we set the
threshold for the extreme response times to 15.7 seconds to fit
the GEV distributions. This threshold lies on the three standard deviations of the mean execution times
(i.e., it is higher than 99.7\% of observed data). 
We use the maximum log-likelihood to estimate the parameters of the GEV distribution
using the observed data. 
The empirical probability density function against the modeled one is shown
in Figure~\ref{fig:SB_density}. We infer the location, scale,
and shape of GEV to be 16.3 (+/- 0.1), 0.4 (+/- 0.1), and 0.0 (+/- 0.1) where
the numbers in the parentheses show the standard errors. 

\vspace{0.5 em}
\noindent \textbf{Validity}.
Since the shape is zero, the type of EVT distribution is I: the tail
is infinite but decays exponentially. Therefore, the extrapolations
can be valid for extrapolation up to a bounded horizon.
Another way to validate the GEV model is to examine the QQ plots,
a plot of various quantiles of the data (empirical)
against the quantiles predicted by the GEV model.
Each point in the plot represents a particular quantile (eg., $5$-{th} percentile) from the data vs. that predicted by the model. An ideal fit is denoted by a $45^{\circ}$ line in the plot. The QQ-plot is shown in Figure~\ref{fig:SB_qq}.
Based on the plot, we can ensure the validity of extrapolation up to an extreme computation
time of 16.8 (s). 

\vspace{0.5 em}
\noindent \textbf{Usefulness}.
Figure~\ref{fig:SB_return_level} shows the m-return level of extreme convergence times of the algorithm. 
For example, the 1000-return level quantifies the expected
extreme execution times (i.e, the expected all timings that exceed the threshold)
that will be observed in the next 1,000 queries.
Since we only have 4,419 response times, to validate the usefulness of the extrapolations,
we use the first 419 traces to infer GEV
distributions and compare the prediction of worst-case convergence times (WCCT) to the
actual observed worst-case WCCT for the next 500, 1000, 2000, and 4000 queries.
The GEV-based extrapolations show the return levels are 16.0 (s) [15.4, 16.7],
16.3 (s) [15.5, 17.0], 16.5 (s) [15.4, 17.7], and 16.9 (s) [15.4, 18.4], respectively,
where the intervals show 95\% lower and upper confidence intervals.
The actual WCCTs are 16.7, 16.7, 16.8, and 17.2 in the next 500, 1000, 2000, and 4000
queries that are well within 95\% confidence intervals.
Furthermore, we can use GEV distributions to calculate the likelihood of extreme execution costs in the next $s$ steps.
The chance of observing extreme response times of 16.8 (s) or more in the next 1,
10, 100, and 1,000 queries are 0.1\%, 1.1\%, 10.7\%, and 77.8\%, respectively.

\section{Probabilistic Worst-Case Convergence Time}
\label{sec:definition}
We study the non-functional behaviors of machine learning software and seek to provide
statistical guarantees on the worst-case convergence time. We consider two stages in the machine learning life cycle:
1) training stage where the ML model is synthesized from the data and hyperparameters using core ML training algorithms; and
2) inference stage where the prediction is made for a given query by the pre-trained ML model.  

\vspace{0.5em}
\noindent{\bf The Convergence of ML.} We define the convergence separately for training
and inference in our setting. 

\noindent \textit{Training Convergence.} We can abstractly view a training stage as the problem of identifying a
mapping $M: \Xx \to \Yy$ from a set $\Xx$ of inputs to a set $\Yy$ of outputs by
learning from a fixed dataset $\Dd = \{({\bf x_i}, {\bf y_i})\}_{i=1}^N$ so that
$M$ generalizes well to previously unseen testing data $\Dd_* = \{({\bf (x^*_i)}, {\bf y^*_i})\}_{i=1}^M$.
In doing so, the training involves configuration parameters---characterizing the set $\Hh$ of hyperparameters---that let the
users define the hypothesis class for the learning tasks. Given a hyperparameter configuration $h \in \Hh$, 
the ML training sift through the given dataset $\Dd$ to learn
an ``optimal'' value $\theta \in \Theta_h$ and thus compute the learning model
$M(\theta|\Dd,h): \Xx \to \Yy$ automatically. In doing so,
the training process can be seen as an optimization algorithm to minimize
a loss function (e.g., impurity, cross-entropy, hinge). Given a criterion such as
the number of iterations in the DNN or tolerance in the support vector machines;
the convergence of the training algorithm is to reach a state that satisfies one of the stopping criteria and returns
the model parameter $\theta$.

\noindent \textit{Inference Convergence.} 
Let $\theta^* \in \Theta_h$ be an optimal parameter (i.e., a model with the minimum
loss) inferred from the training stage. Let $M_{\theta^*}: \Xx \to \Yy$ be the corresponding
model to infer a prediction $y \in \Yy$ for a query $x \in \Xx$.  The convergence time
is the time taken to derive predictions by the model and reach a desirable goal (e.g.,
for a model that controls a robot; it is the time taken to derive the robot from
its initial state to a final state by predicting the next action in each step). 

\vspace{0.5em}
\noindent{\bf Convergence Time.}
We use a high-level cost model to define the convergence time.
The cost model of a ML $M$ is the (non-functional) cost of computations
(e.g., running time, memory usages, network packets, etc).
Rather than abstracting costs with notions such as the number of byte-code
executed, we use the actual execution times. The convergence time is thus the computation cost of
reaching to the convergence state from an initial state during the training or inference stages. 

In our monitoring setting, we observe the queries of ML over different input values $\{x_1,\ldots,x_n\}$ and record the
convergence times $\{t_1,\ldots,t_n\}$.
Since we potentially have a large number of samples, we can compute both the expected values
and the standard deviation of costs and thus, conclude
whether the difference of the mean value from 0 is statistically significant.
For example, if a value exceeds two standard
deviations, then this value is statistically significant in the convergence time with
certainty 95\%. 

From a practical standpoint, it is crucial to ensure that there are no significant deviations
from the mean, but also to model and reason about the cost of ML training and inference on the
worst-case scenarios (i.e., the tail of convergence times) such as the expected
worst-case convergence in the future. 

\begin{definition}[Problem Statement]
    \label{def-problem}
    % \begin{tcolorbox}[boxrule=1pt,left=1pt,right=1pt,top=1pt,bottom=1pt]
     Given an ML model $M$ and a set of input traces $\{x_1,\ldots,x_n\}$, our goal is to (1) model the statistics of worst-case convergence times, i.e., $M_n = \max_{i=1}^{n} t_i$; (2) study the validity of extrapolation to provide probabilistic guarantees on the worst-case convergence; and (3) quantify the frequency and significance of extreme convergence time in a long horizon.
\end{definition}

We are interested in the maximum value of convergence times, denoted as $M_n$, over a large population of $n$ samples. 
Each sample is a random variable that differs from one sample to another. 
Since we assume that the values, corresponding to different samples, do not depend on each other, the largest value $M_n$ can be viewed as the maximum of a large number of independent identically distributed random variables.
Therefore,  extreme value theory (EVT) is an ideal framework to study the corresponding limit distributions and the convergence to these distributions. We adapt the statistics of EVT to model, predict, and quantify the worst-case convergence time of ML algorithms for both training and inference stages.

\section{Experiments}
\label{sec:def}
We first provide details on the implementations. Then, we discuss the scalability and validity of EVT
on a set of classic textbook algorithms implemented in Python
and run in a super-computing machine. Finally, we show the feasibility,
scalability, and usefulness of EVT on bounding the worst-case convergence times of ML training
and inference. 
% The tool, benchmarks, and source code are available in the following GitHub repository:
% \url{https://github.com/cuplv/WCET\_via\_EVT}.

\subsection{Implementation Details and Research Questions}
We monitor the applications on a super-computing machine with
the Linux Red Hat 7 OS and an Intel Haswell 2.5 GHz CPU with
24 cores (each with 128 GB of RAM). In doing so, we simulate
the queries with random test cases generated independently
and uniformly from the domain of variables.
For training algorithms, we generate the input traces using
\textsc{DPFuzz}~\cite{tizpaz2020detecting},
the state-of-the-art fuzz testing method to
characterize the worst-case performance of ML training algorithms.
For deep neural networks, we simulate the traces
using state-of-the-art methods for the functional
properties of cyber-physical systems~\cite{Dutta+Others/2018/Learning}.
We implemented the EVT algorithms in \texttt{R} using \texttt{evd}
and \texttt{extRemes} libraries~\cite{gilleland2013software}.
We used \texttt{mrlplot()} to pick thresholds, \texttt{fevd()} to
fit extreme value distribution, and \texttt{return.level()} to extract
return levels.

\begin{table*}[tbp!]
    \caption{Convergence Times of Training Linear Support Vector Machine with Prevalent Datasets of Various Sizes.
    Since the Bayesian factor (BF) is 95\% confidence on the WCCT after observing $b$ samples, we only use $b$ samples to
    infer GEV distributions.
      Legend: \textbf{\#N}: Size of Training Dataset, 
      $\overline{\textbf{T}}$: Average of Convergence Time (ms),
      \textbf{T}$_n$: Actual Max. Convergence Times Observed after $n$ Queries (ms),
      $b$: The number of samples observed until the Bayes factor (BF) convinced,
      \textbf{T}$_b$: Actual Observed WCCT up to the $b$-th sample 
      (this is the prediction of Bayes factor),
      \textbf{Parameters}: ($\mu, \sigma, \xi$) of GEV distribution,
      \textbf{$\tau$}: Thresholds of GEV distribution (s),
      \textbf{RL$_n$}: GEV-based Prediction of Max. Convergence Times after $n$ Queries (ms),
      \textbf{Error}$_n$: $\frac{RL_n - T_n}{T_n - T_b}$: the percentage of Error of GEV predictions,
      compared to the baseline Bayes factor (an error of -1.0). Note: $K=10^3$.
    }
    \label{tab:EVT-benchmark}
    \resizebox{\textwidth}{!}{
      \begin{tabular}{ | l  l | l  l  l  l  l | l  l | l  l  l  l | l  l  l  l  | l  l  l |}
      \hline
     \multicolumn{2}{|c|}{Benchmark}  &   \multicolumn{5}{c|}{Observed Convergence Time} & \multicolumn{2}{c|}{BF} & \multicolumn{4}{c|}{GEV Characteristics} & \multicolumn{4}{c|}{GEV-based Predictions} & \multicolumn{3}{c|}{Accuracy} \\  \hline %\cline{4-17}
    Dataset & \textbf{\#N} & $\overline{\textbf{T}}$ & \textbf{T}$_{1K}$ & \textbf{T}$_{2K}$ & \textbf{T}$_{5K}$ & \textbf{T}$_{10K}$ & $b$ & \textbf{T}$_{b}$ &  $\mu$ & $\sigma$ & $\xi$ & \textbf{$\tau$} & \textbf{RL}$_{1K}$ & \textbf{RL}$_{2K}$ & \textbf{RL}$_{5K}$ & \textbf{RL}$_{10K}$ & \textbf{Error}$_{1K}$ & \textbf{Error}$_{5K}$ & \textbf{Error}$_{10K}$ \\ \hline
    % $\frac{410-538}{538 - 401}$, $\frac{462-538}{538 - 401}$, $\frac{489-538}{538 - 401}$
    Census & 6512 & 169.3 & 538 & 538 & 538 & 538 & 148 & 401 & 340 & 18.7 & 0.0 & 250 & 410 & 431 & 462 & 489 & -0.93 & -0.38 & -0.36 \\ \hline
    Census & 13024 & 323.7 & 630 & 712 & 778 & 844 & 261 & 555 & 584 & 17.3 & 0.0 & 520 & 644 & 666 & 697 & 724 & +0.19 & -0.36 & -0.41 \\ \hline
    Census & 19536 & 480.7 & 1235 & 1235 & 1235 & 1235 & 101 & 1117 & 956 & 46.5 & 0.0 & 757 & 1112 & 1175 & 1263 & 1301 & -1.05 & +0.24 & +0.56 \\ \hline
    Census & 26048 & 681.5 & 1780 & 1816 & 1966 & 2147 & 244 & 1553 & 1312 & 47.2 & 0.1 & 1010 & 1671 & 1830 & 2082 & 2311 & -0.48 & +0.28 & +0.27 \\ \hline
    Census & 32235 & 827.3 & 1990 & 2504 & 2504 & 2673 & 140 & 1670 & 1673 & 92.1 & 0.0 & 1246 & 2054 & 2178 & 2339 & 2463 & +0.20 & -0.20 &  -0.21 \\ \hline \hline
    Bank & 9042 & 392.1 & 636 & 636 & 654 & 654 & 134 & 636 & 492 & 23.1 & 0.0 & 392 & 553 & 577 & 607 & 632 & -9.99 & -2.10 & -1.20 \\ \hline
    Bank & 18084 & 477 & 1249 & 1249 & 1283 & 1357 & 168 & 819 & 962 & 31.9 & 0.1 & 747 & 1093 & 1161 & 1265 & 1354 & -0.36 & -0.04 & -0.01 \\ \hline
    Bank & 27126 & 740 & 2089 & 2089 & 2319 & 2515 & 94 & 1276 & 1453 & 58.9 & 0.1 & 1111 & 1814 & 1956 & 2171 & 2358 & -0.34 & -0.14 & -0.13 \\ \hline
    Bank & 36168 & 1069 & 2822 & 3703 & 3703 & 3772 & 94 & 1862 & 2240 & 140.1 & 0.0 & 1570 & 2765 & 2939 & 3186 & 3384 & -0.06 & -0.28 & -0.20 \\ \hline
    Bank & 44758 & 1322 & 3515 & 4378 & 4378 & 4378 & 193 & 3515 & 3096 & 220.6 & 0.0 & 2066 & 3931 & 4212 & 4612 & 4935 & -9.99 & +0.27 & +0.65 \\ \hline
    \end{tabular}
 }
 % \vspace{-0.5em}
\end{table*}

\begin{enumerate}[start=1,label={\bfseries RQ\arabic*},leftmargin=3em]

\item How do GEV-based predictions of WCCT compare to the baseline statistical
testing via Bayes factor on the set of linear training algorithms? 

\item How accurate and useful are GEV-based extrapolations of WCCT
of training popular ML algorithms? 

\item How accurate and useful are GEV-based extrapolations for
the inference computation times of DNN models? 

\end{enumerate}

\subsection*{RQ1: Accuracy of GEV compared to the Bayes factor method.}
We perform experiments on $10$ benchmarks over the linear support vector
machines. We use two classical datasets: census~\cite{Dua:2019-census}
is a binary classification dataset that predicts whether an individual has an income over $50K$ a year.
The dataset has $14$ attributes. 
Bank Marketing~\cite{Dua:2019-bank} is another classic tabular dataset
that predicts whether an individual, described with $17$ features,
subscribes to the term deposit of the bank.
We chose the linear model with these datesets as the baseline ML
algorithms to evaluate the efficacy of GEV, in comparison to the
prevalent statistical testing. Specifically, we consider
Jeffreys test~\cite{jha2009bayesian,sankaranarayanan2013static}, a variant of Bayes factor,
with a uniform prior that finds a lower-bound on the number of successive samples $K$ that is sufficient for us to
accept a maximum observed computation time as the WCCT of the training algorithms.
The number of such samples $K$ is obtained via:
\[
K \geq \lceil(-\log_2 B)/(\log_2 \theta)\rceil  
\]
where $B$ in the numerator is Bayes factor and can be set to $100$ for a very strong
evidence. For instance, to achieve a $\theta = 0.95$ (a confidence of 95\%), we are required to
set $K \geq 90$ to be highly confident in accepting the worst-case computation times.

Table~\ref{tab:EVT-benchmark} shows the performance of GEV extrapolations
compared to the baseline Bayes factor. We use 20\%, 40\%, 60\%, 80\%,
and 95\% of Census and Bank for $5$ different training scenarios per each
dataset. We report observed the average and worst-case convergence times
of training (\textbf{T}), the number of samples to convince the Jeffreys test with the
95\% confidence (\textbf{b}), the prediction based on the test (\textbf{T}$_b$),
the characteristics of GEV distributions inferred using \textbf{b} samples
when the test convinced, the predictions based on GEV distributions (\textbf{RL}),
and the error of GEV prediction as compared to the baseline.

To measure the prediction error, we compare the GEV prediction of the
$n$-th queries (\textbf{RL}$_n$) to the actual WCCT at the $n$-th queries
(\textbf{T}$_n$), while factoring the baseline Jeffreys test prediction (\textbf{T}$_b$)
as following: $\frac{RL_n - T_n}{T_n - T_b}$. Negative values show that the GEV-based
prediction under-estimates the actual WCCT whereas positive values show that
the GEV-based prediction over-estimates the actual WCCT. The absolute value shows
the percentage of error. For example, in Table~\ref{tab:EVT-benchmark} for the Census dataset with 32235 samples,
the GEV used only 140 initial samples and predicted the WCCT in the next 10,000
queries with an error of 21\% (under-estimated).  Note that GEV and Jeffreys have the same error rate
when the error is +1.0 or -1.0 whereas values between -1.0 and +1.0 show GEV outperformed 
Jeffreys test and value below -1.0 or above +1.0 show Jeffreys achieved better
results. We truncated any error below -10.0 or above +10.0.  

Overall, Table~\ref{tab:EVT-benchmark} shows that GEV-based predictions
have lower prediction errors compared to the baseline Jeffreys test
in 25 cases out of 30 cases whereas the Jeffreys test is more accurate in 5
cases (cases with errors less than -1.0 such as Bank with 9042 data samples). 
Since Jeffreys test accepts the WCCT after at most 261 samples (see the column
$b$); the GEV prediction used at most 261 samples to extrapolate for the
next 1K, 2k, 5k, and 10k queries (the column $b$ shows the number of samples
used to derive GEV parameters). In 22 cases out of 30 cases;
GEV prediction under-estimates the actual WCCT whereas in the remaining
8 cases, it overestimates the actual WCCT. The errors of GEV predictions
are below 50\%, 25\%, and 10\% in 22, 11, and 3 cases out of 30, compared
to the Jeffreys test.

\begin{tcolorbox}[colback=aliceblue, boxrule=1pt,left=1pt,right=1pt,top=1pt,bottom=1pt]
\textbf{Answer RQ1:}
Overall, GEV predicts the worst-case convergence times more accurately than the baseline
Bayes factor method in 83\% of cases. Compared to the baseline, GEV predictions are more than 50\%,
75\%, and 90\% accurate in 73\%, 37\%, and 10\% of cases, respectively.  
\end{tcolorbox}

\subsection*{RQ2: GEV predictions of WCCT for ML training algorithm.}
\label{sec:case-stuy}
We consider the application of EVT for bounding the worst-case
convergence times in training classical (non-neuron) ML models.
Our goal is to evaluate the \textbf{feasibility}, \textbf{usefulness},
and \textbf{scalability} of extreme value theory in modeling, quantifying,
and bounding the worst-case convergence times of ML training phase. 
In doing so, we consider the training of four mature ML training
algorithms and wish to provide an upper-bound on the convergence time of training via extreme value theory.
For each case study, we discuss whether such a bound is possible,
the frequency of extreme training convergence time, and the expected return levels.

Rather than using the Jeffreys test~\cite{jha2009bayesian}, we follow the standard methods in selecting
the threshold of extreme value to fit GEV distributions, according to Coles et al.~\cite{coles2001introduction}.
The initial threshold is set to the mean of
samples plus two standard deviations and implies that only 4.56\% of data samples are considered as the tail
samples. Since this might not give us a valid GEV (examples of invalid GEV include 
a shape of 1.0 or more, negative return levels, and decreasing return levels as the number of queries
increases), we vary the threshold down to the mean of samples plus one standard deviation (with a rate of 0.05\%).
If we could not find any valid GEV, we return failed. We say a GEV prediction is accurate if
it includes the actual WCCT within its 95\% confidence range of prediction. 

%  Cross-out for now: bring it back
\begin{figure*}[tbp!]
    \centering
    \begin{subfigure}[b]{0.24\textwidth}
        \includegraphics[width=\textwidth]{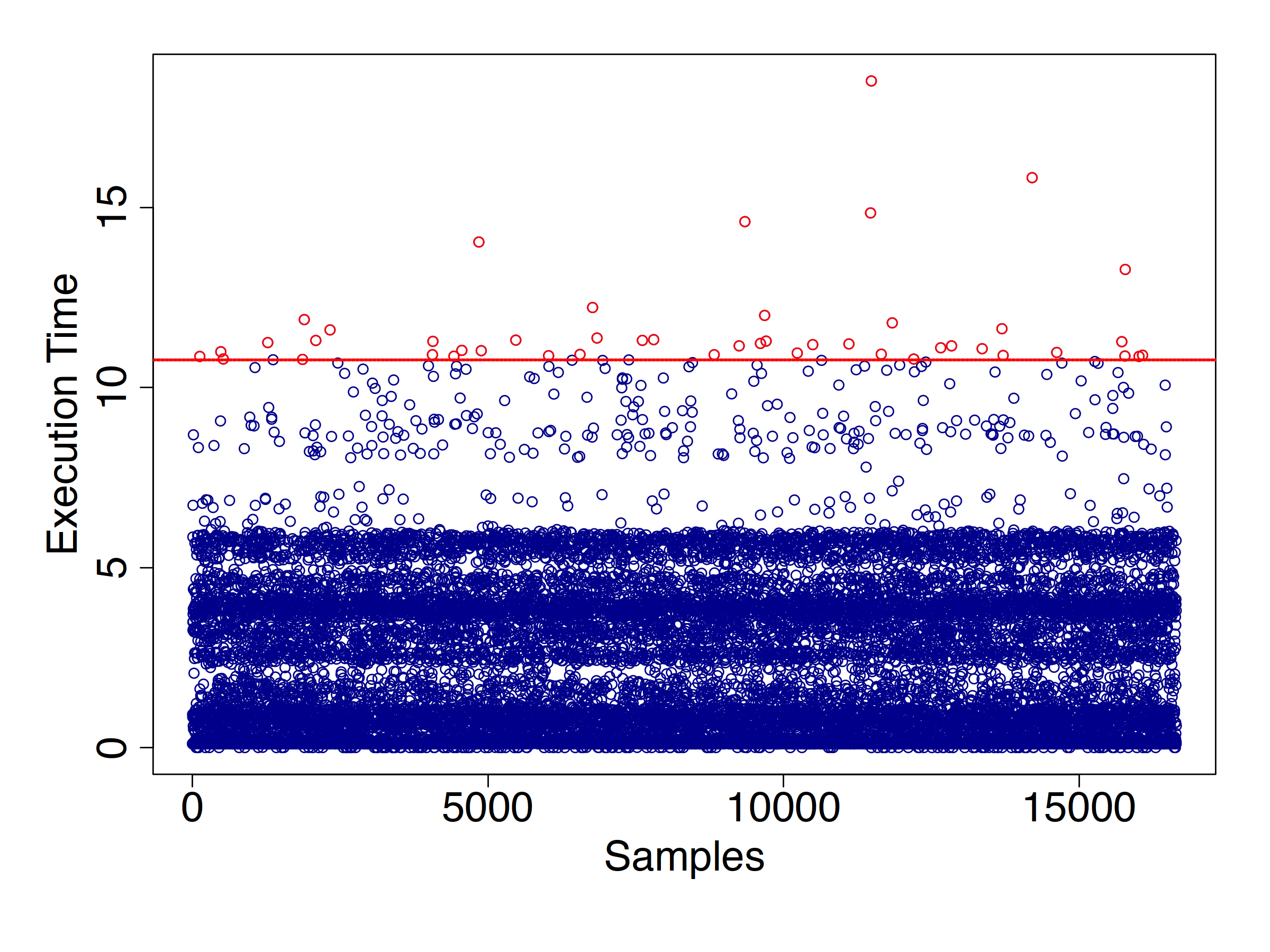}
        \caption{}
        \label{fig:LR_threshold}
    \end{subfigure}
    \begin{subfigure}[b]{0.24\textwidth}
        \includegraphics[width=\textwidth]{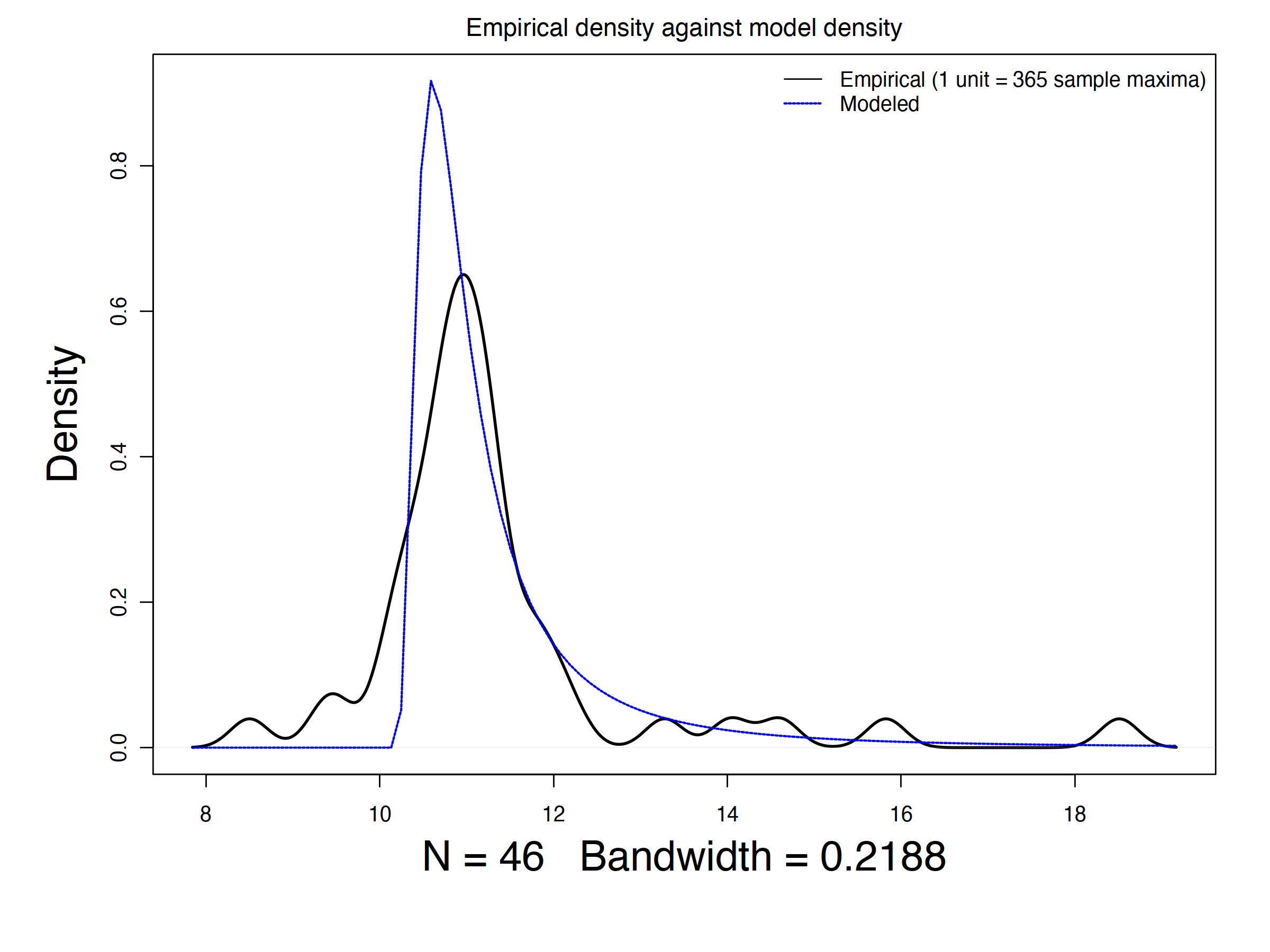}
        \caption{}
        \label{fig:LR_density}
    \end{subfigure}
    \begin{subfigure}[b]{0.24\textwidth}
        \includegraphics[width=\textwidth]{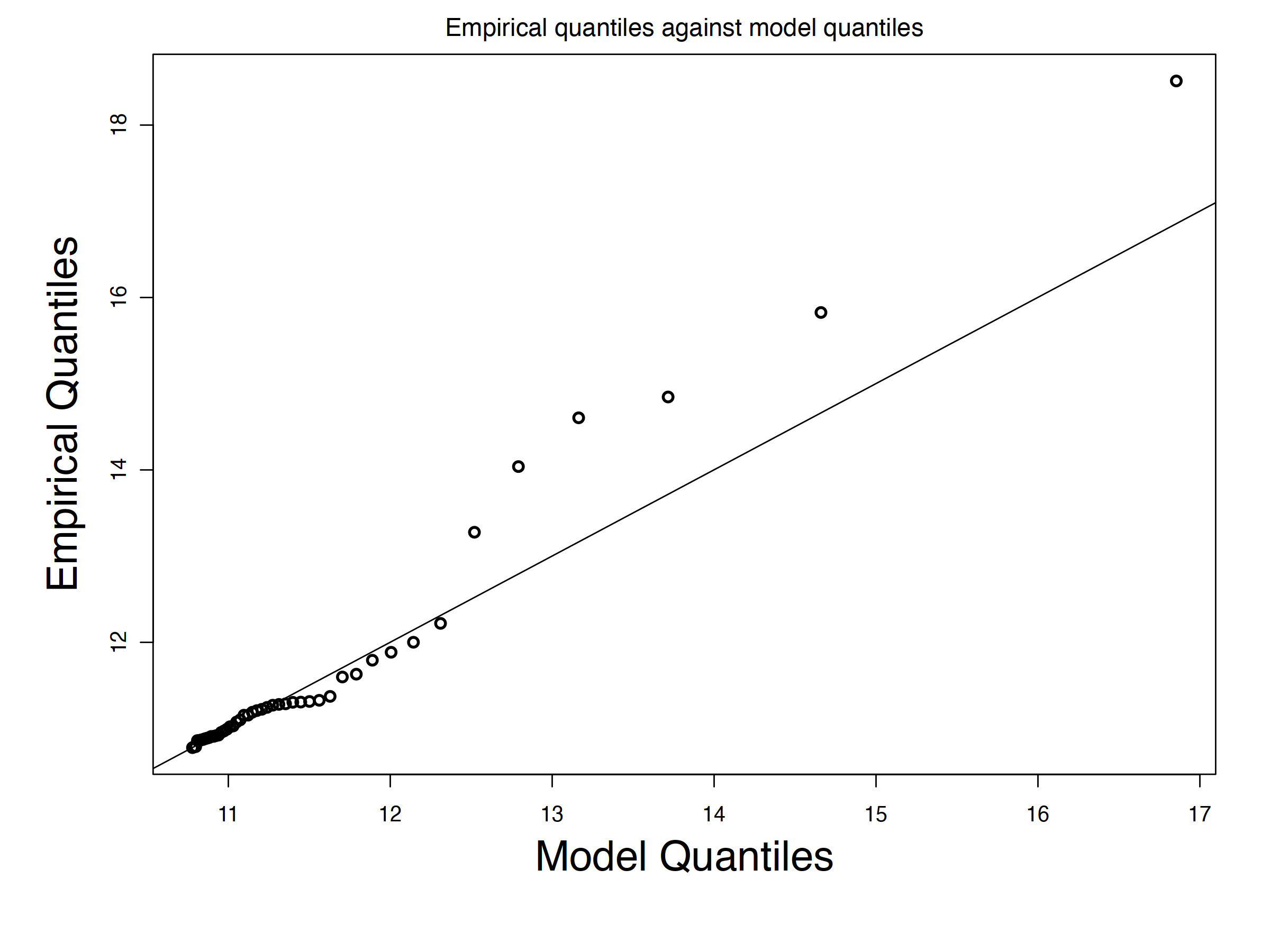}
        \caption{}
        \label{fig:LR_qq}
    \end{subfigure}
    \begin{subfigure}[b]{0.24\textwidth}
        \includegraphics[width=\textwidth]{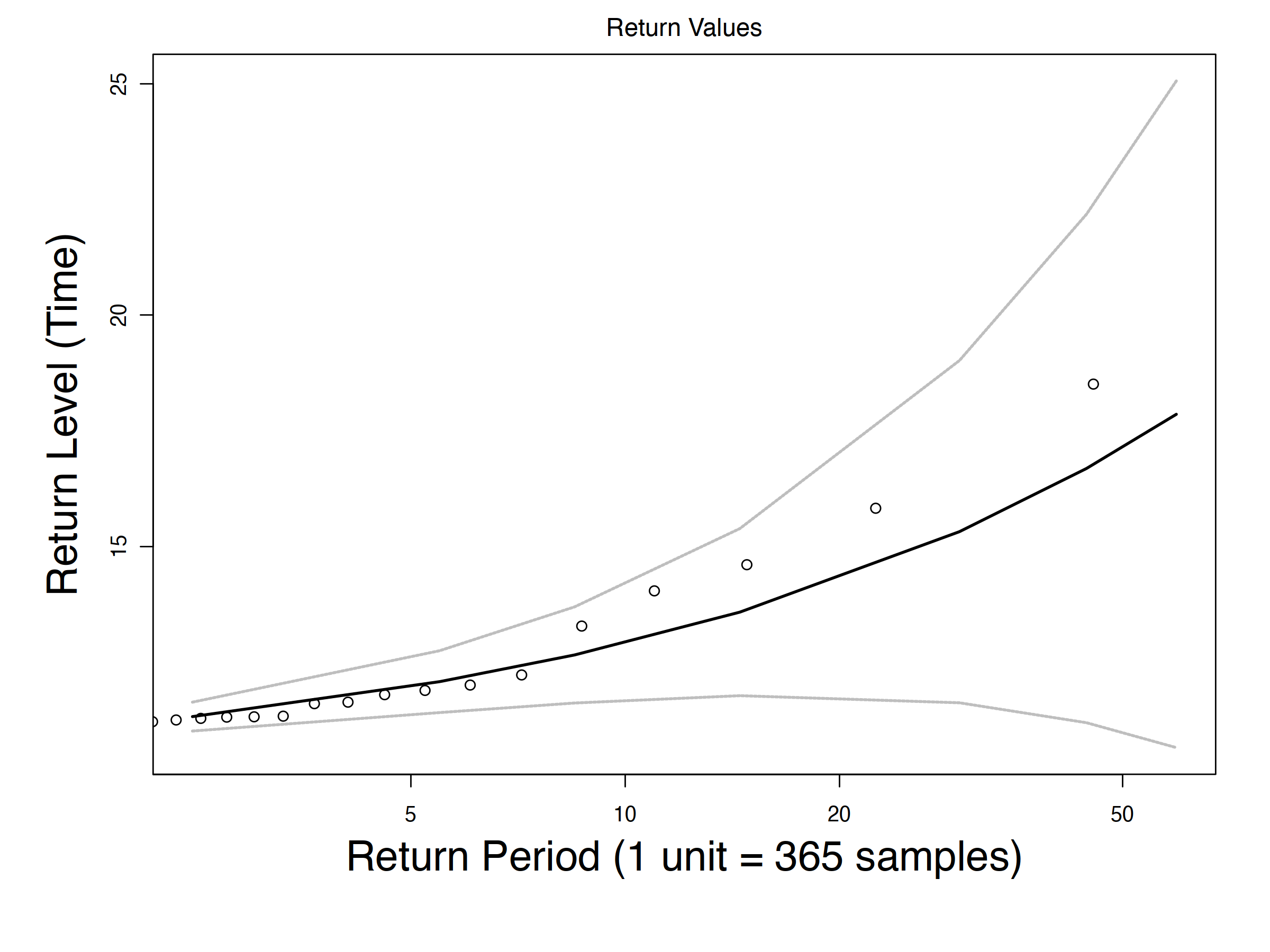}
        \caption{}
        \label{fig:LR_return_level}
    \end{subfigure}
    \caption{Logistic Regression. (a) training convergence times for logistic regression varying input dataset and hyperparameters where the red line shows 0.997-quantile (99.7\% of data is below the red line),
    (b) the density plot for GEV of Logistic Regression,
    (c) quantile plot for execution time of Logistic Regression,
    (d) m-return level plot for Logistic with expected values and their 95\% CI.
}
\end{figure*}

\vspace{0.5 em}
\noindent \textbf{Logistic Regression.} 
Logistic regression is a popular classifier that supports
various linear and non-linear solvers such as newton and saga. We
study the implementations in scikit-learn library~\cite{scikit-learn}.

\vspace{0.25 em}
\noindent \textit{Test Cases.} We used \textsc{DPFuzz}~\cite{tizpaz2020detecting}
with i.i.d mode and randomly generate the training dataset
via the dataset synthesizer library~\cite{make-multilabel-classification}. 
Within $4$ hours, we generate $16,568$ models with different
training datasets as well as hyperparameters. The computation times of
these training tasks are inputs to the EVT analysis. 

\vspace{0.25 em}
\noindent \textit{Feasibility and Scalability.} Figure~\ref{fig:LR_threshold}
shows the recorded computation times as well as the threshold of 10.7 for extreme values.
It takes less than $2$ seconds to infer GEV distributions. Figure~\ref{fig:LR_density} shows
the empirical vs. modeled probability density functions. The location, scale, and
shape of GEV are 10.8, 0.49, and 0.9, respectively. Figure~\ref{fig:LR_qq} shows the QQ plot.

\vspace{0.25 em}
\noindent \textit{Usefulness.} The return levels are shown in Figure~\ref{fig:LR_return_level}.
For 500, 1K, 2K, 5K, and 10K queries, the predicted return levels are 11.0 (s) [10.9, 11.4],
11.4 (s) [10.9, 11.8], 12.6 (s) [11.4, 13.7], 15.9 (s) [11.9, 19.9], and 20.4 (s) [10.6, 30.2], respectively.
The actual observed times are 14.8 (s), 26.8 (s), 26.8 (s), 26.8 (s), and 26.8 (s), respectively,
for 500 to 10K queries. While the 95\% confidence range predictions do not include the
actual WCCT up to 5K queries; the actual WCCT of 10K queries are within the range of
predicted WCCT. Based on the GEV distributions, we expected to observe an extreme convergence time of
11.0 (s) in every 793 queries and the likelihood of observing such event next is 0.12\%.

\vspace{0.5 em}
\noindent \textbf{Decision Tree.} 
The decision tree classifier is a popular white-box classifier that partitions the space of
training data into hyper-rectangular sub-spaces to learn predicates for each class label.
We analyze its implementations in scikit-learn library~\cite{scikit-learn}.

\vspace{0.25 em}
\noindent \textit{Test Cases.} Similar to the previous case studies,
we train $9,348$ models within $4$ hours, and our goal is to
estimate the worst-case training convergence time via EVT.

\begin{figure*}[tbp!]
    \centering
    \begin{subfigure}[b]{0.24\textwidth}
        \includegraphics[width=\textwidth]{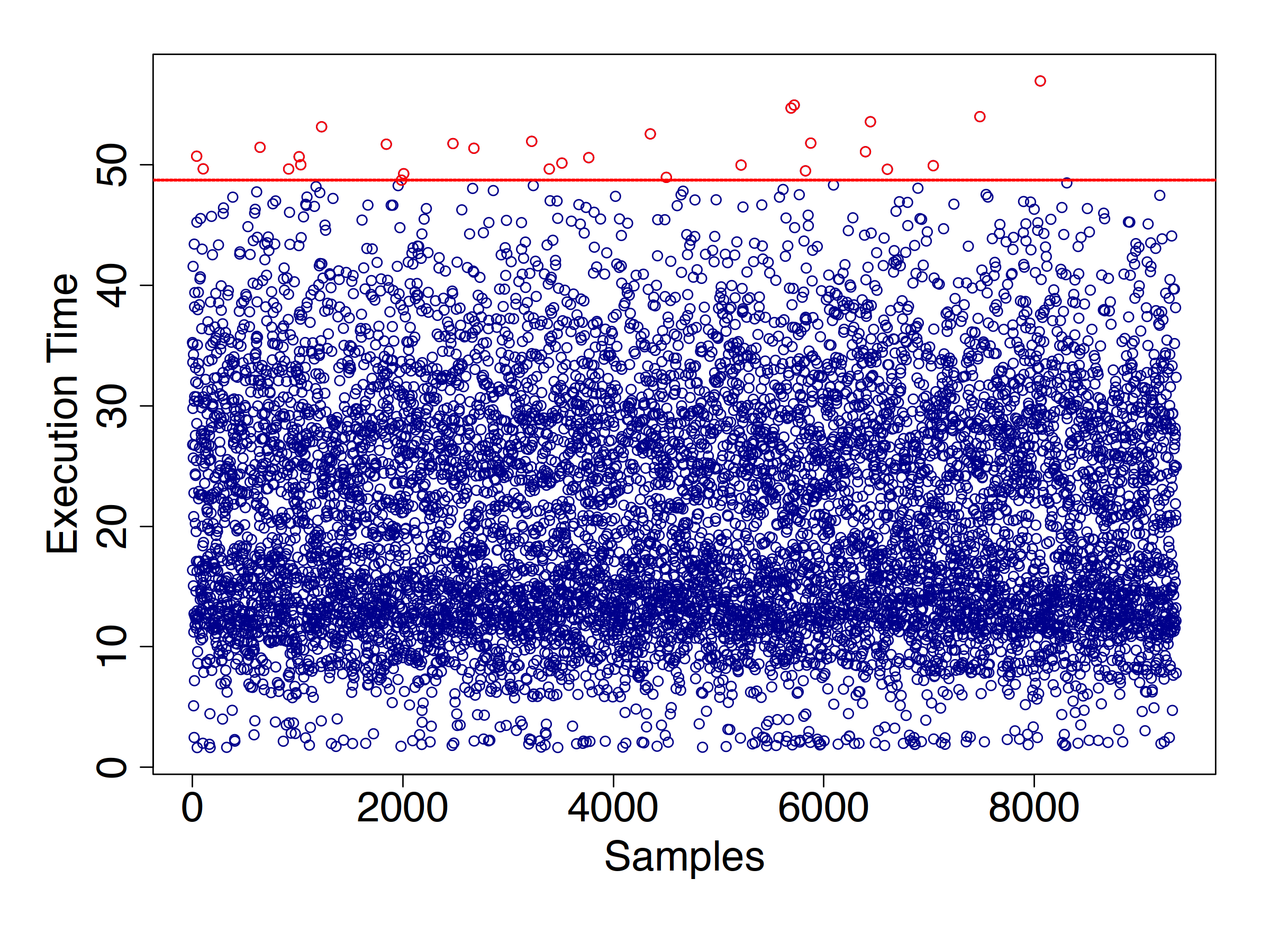}
        \caption{}
        \label{fig:DT_threshold}
    \end{subfigure}
    \begin{subfigure}[b]{0.24\textwidth}
        \includegraphics[width=\textwidth]{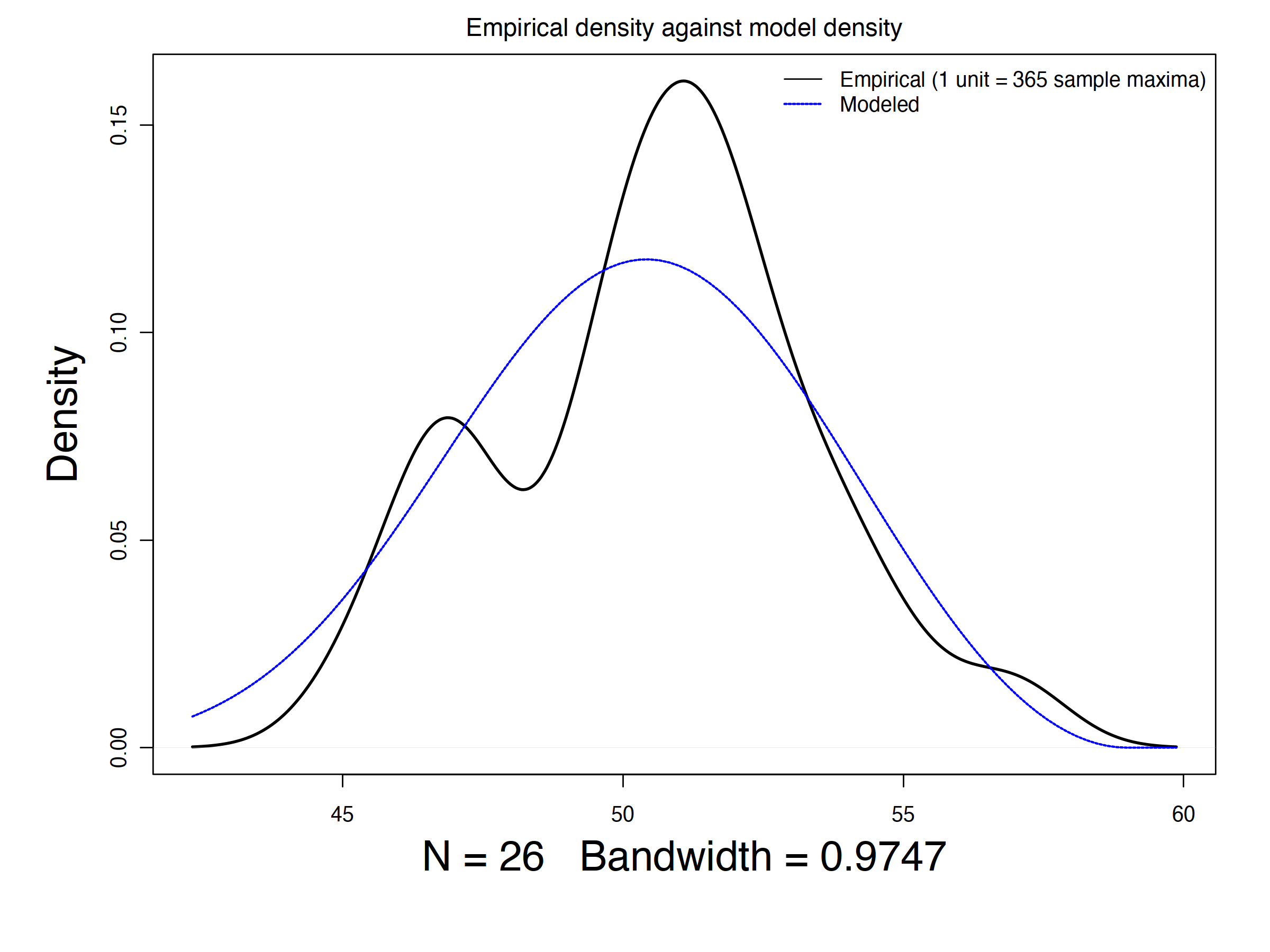}
        \caption{}
        \label{fig:DT_density}
    \end{subfigure}
    \begin{subfigure}[b]{0.24\textwidth}
        \includegraphics[width=\textwidth]{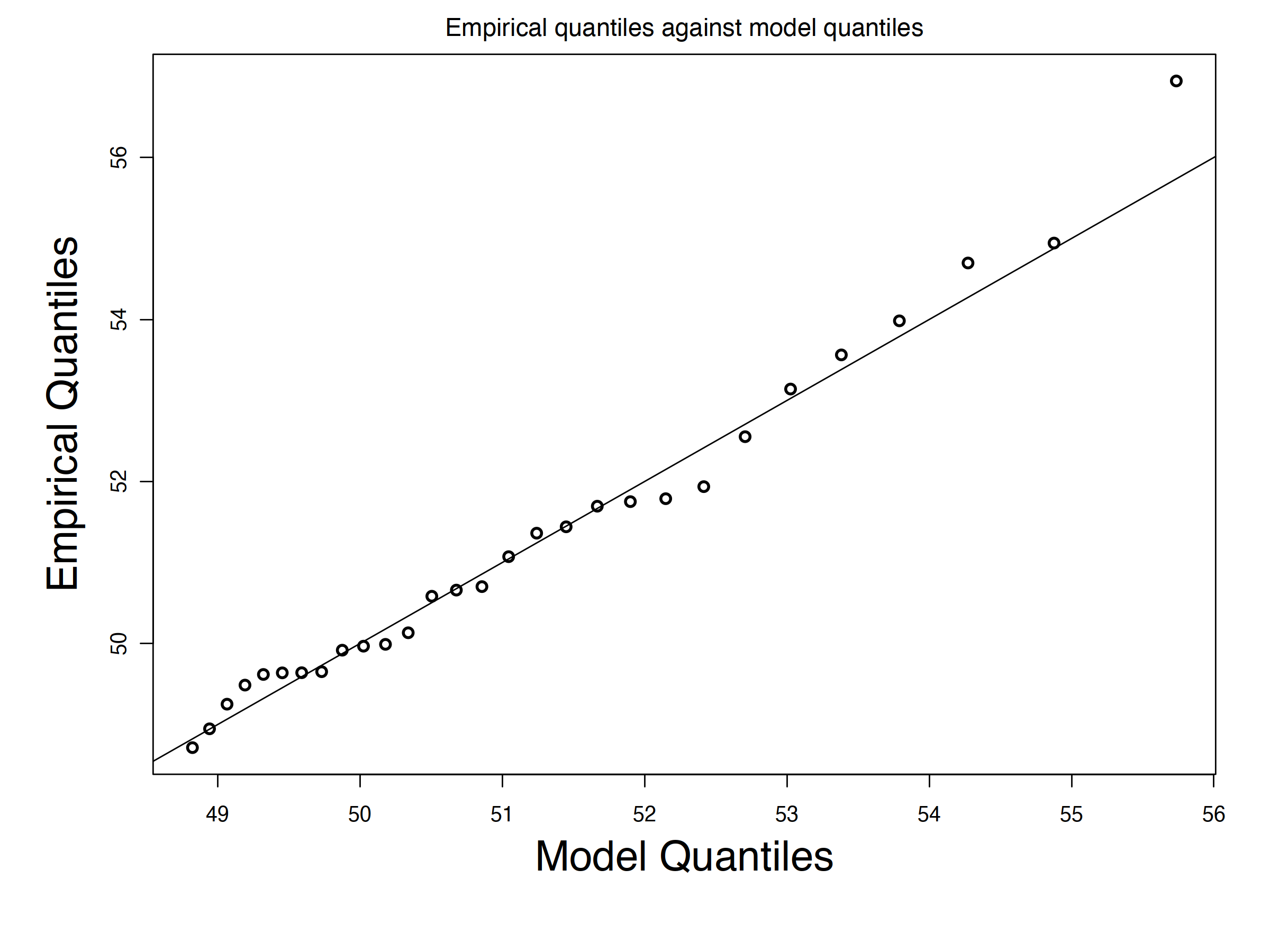}
        \caption{}
        \label{fig:DT_qq}
    \end{subfigure}
    \begin{subfigure}[b]{0.24\textwidth}
        \includegraphics[width=\textwidth]{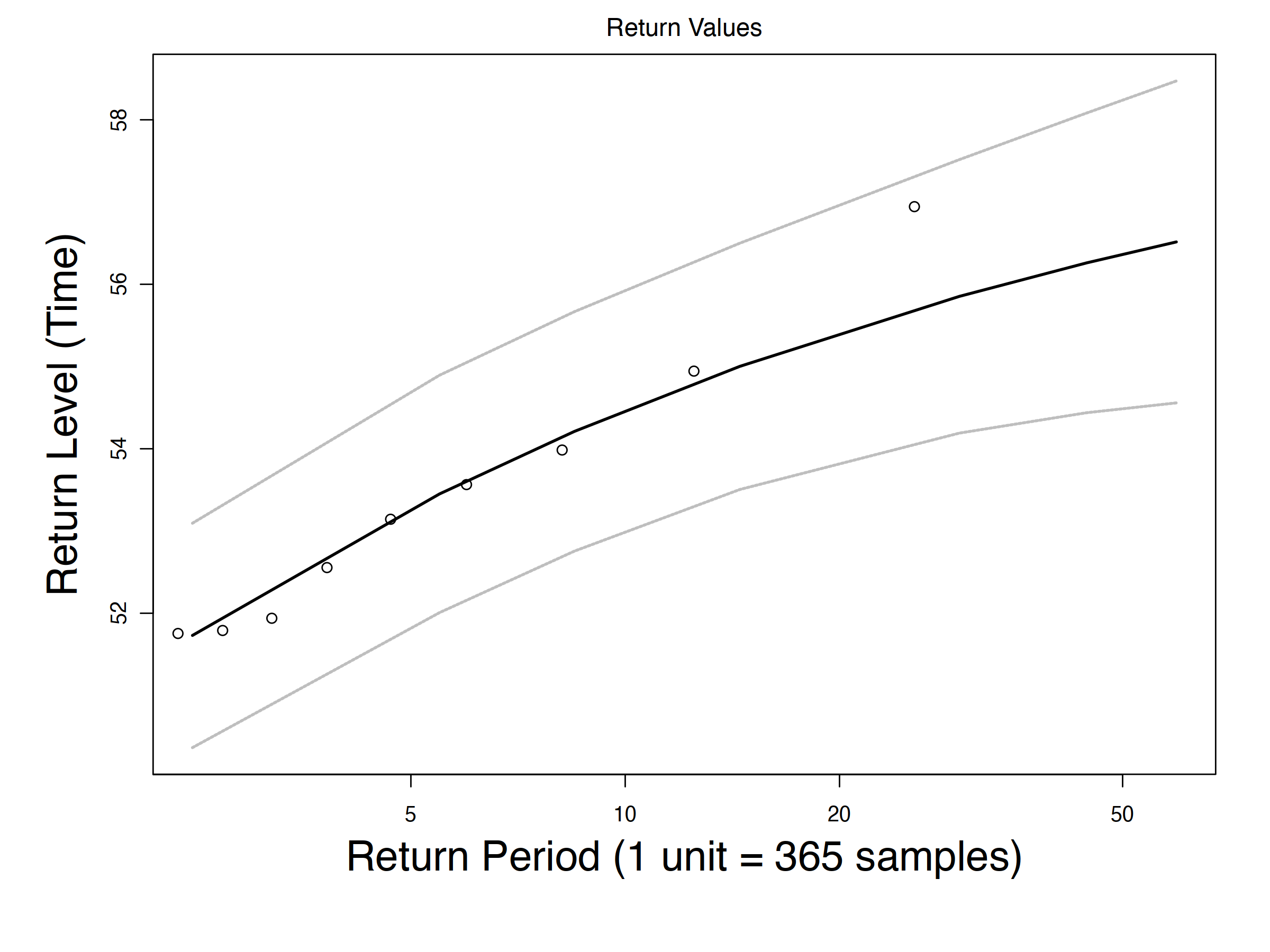}
        \caption{}
        \label{fig:DT_return_level}
    \end{subfigure}
    \caption{Decision Tree. (a) the computation times of training Decision Tree varying input dataset and hyperparameters with a threshold sets to 48.7 (s),
    (b) the density plot for GEV of Decision Tree,
    (c) quantile Plot for execution time of Decision Tree,
    (d) m-return level plot for Decision Tree with expected values and their 95\% CI.}
\end{figure*}

\vspace{0.25 em}
\noindent \textit{Feasibility and Scalability.} Figure~\ref{fig:DT_threshold}
shows the recorded execution times of training and the threshold for extreme
training computation times, which is set to 48.7 (s). Figure~\ref{fig:DT_density} shows
the empirical vs. modeled probability density functions. The location, scale, and
shape of GEV are 49.1, 0.77, and 0.16, respectively. Figure~\ref{fig:LR_qq} shows the QQ plot
that is near-linear.

\vspace{0.25 em}
\noindent \textit{Usefulness.} Figure~\ref{fig:DT_return_level} shows the return levels
and its 95\% confidence intervals. For 500, 1K, 2K, 5K, and 10K return periods,
the return levels are 50.3 (s) [49.2, 51.7], 51.4 (s) [50.1, 52.8], 53.3 (s) [51.8, 54.7],
54.9 (s) [53.4, 56.4], and 55.8 (s) [54.1, 57.4], respectively.     
The actual WCCTs for 500, 1K, 2K, 5K, and 9,348 are 51.4, 51.4, 54.9, 56.9, and 56.9. All
the actual WCCTs are within 95\% confidence intervals of GEV-based predictions.
Through GEV analysis, we also expect to observe an extreme training convergence time of 50.3 (s)
one in every 850 queries, and the likelihood of observing such an extreme training time is 0.1\%.

%  Cross-out for now: bring it back

\vspace{0.5 em}
\noindent \textbf{Linear Discriminant Analysis.} 
The goal of LDA is to find a linear combination of features that maximally separates or discriminates
between different classes. We analyze its implementations in scikit-learn library~\cite{scikit-learn}.

\vspace{0.25 em}
\noindent \textit{Test Cases.} Similar to the previous case studies,
we train $3,900$ models within $4$ hours, and our goal is to estimate the worst-case training convergence time via EVT.

\begin{figure*}[tbp!]
    \centering
    \begin{subfigure}[b]{0.24\textwidth}
        \includegraphics[width=\textwidth]{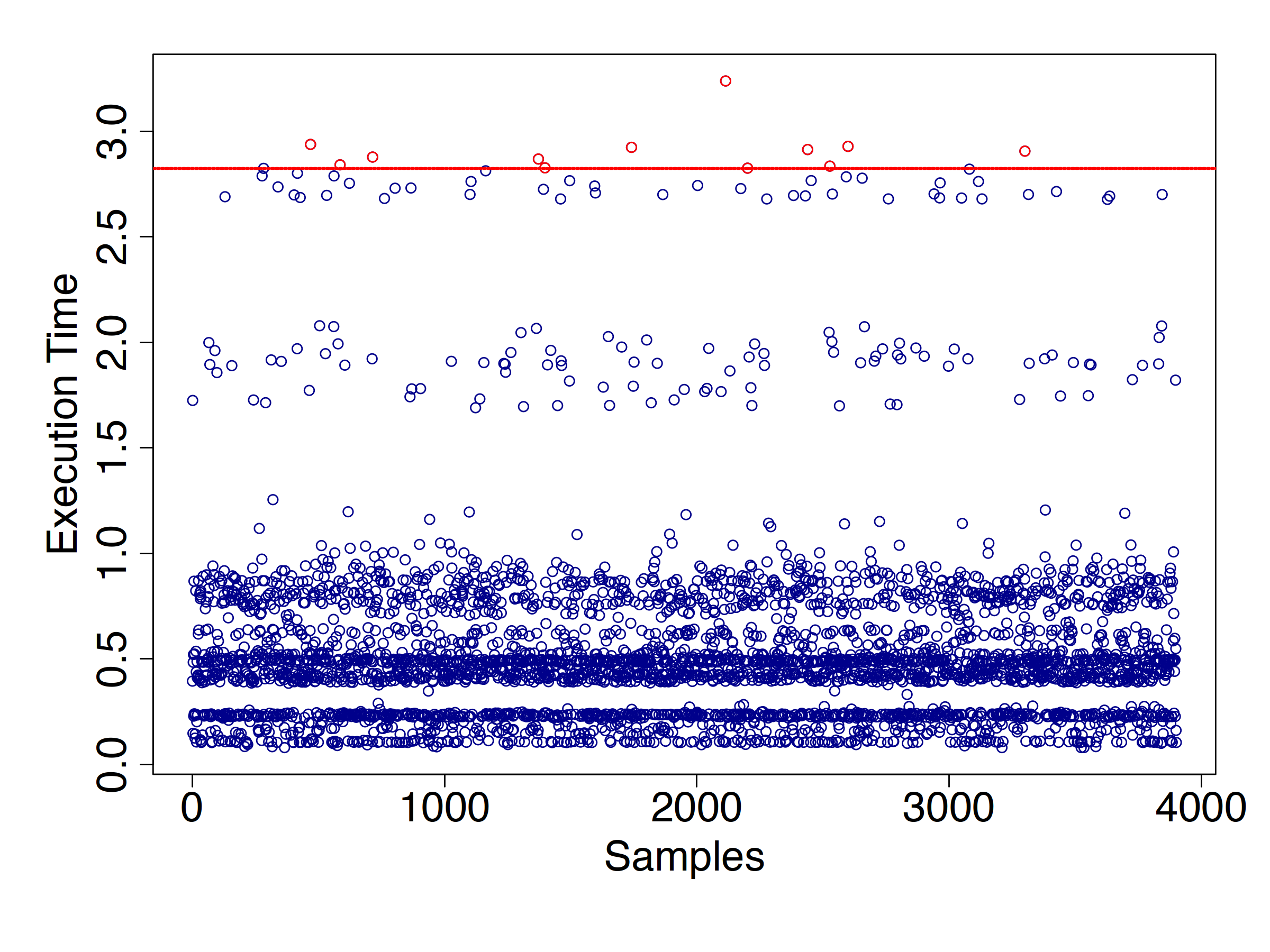}
        \caption{}
        \label{fig:DA_threshold}
    \end{subfigure}
    \begin{subfigure}[b]{0.24\textwidth}
        \includegraphics[width=\textwidth]{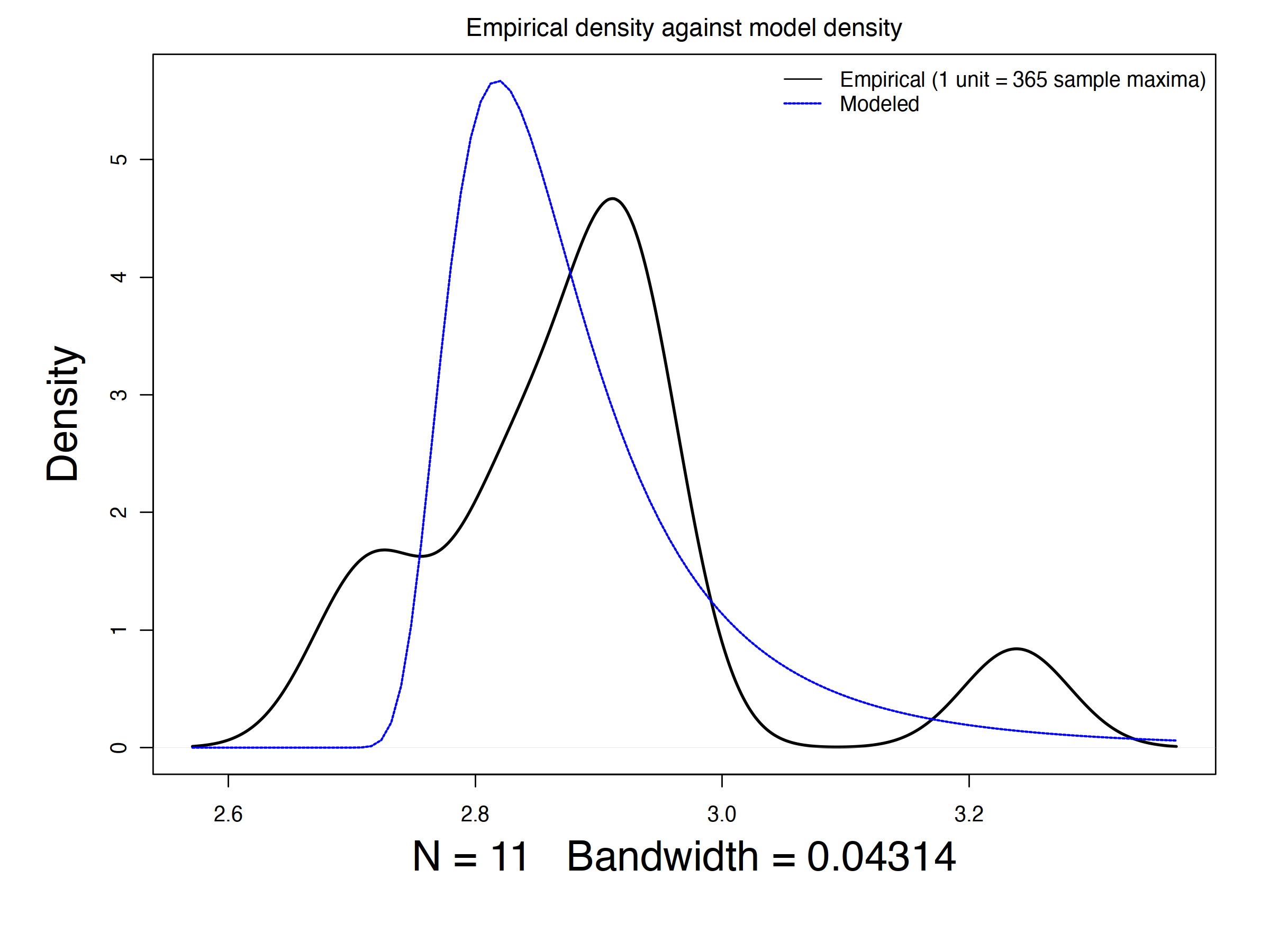}
        \caption{}
        \label{fig:DA_density}
    \end{subfigure}
    \begin{subfigure}[b]{0.24\textwidth}
        \includegraphics[width=\textwidth]{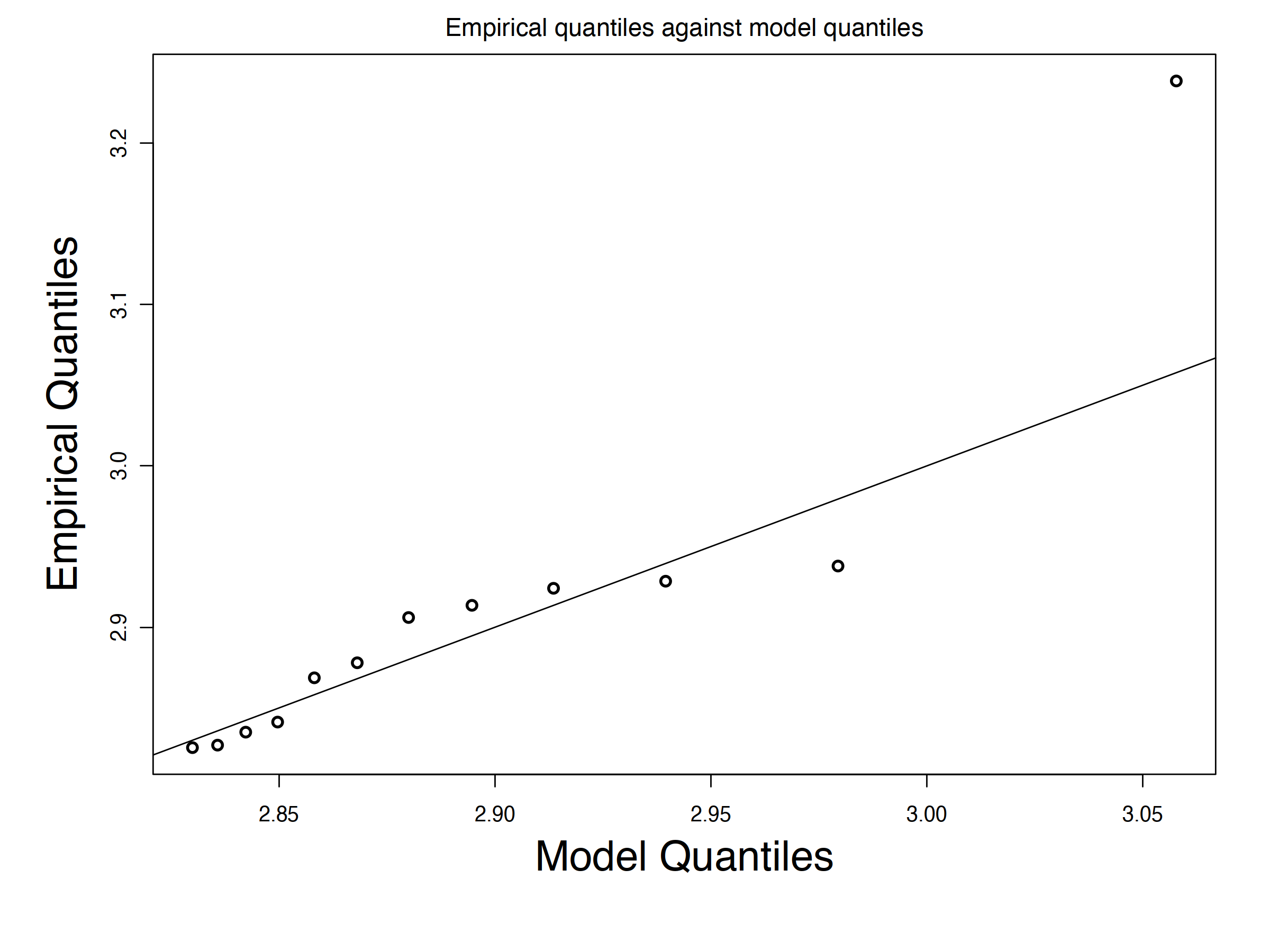}
        \caption{}
        \label{fig:DA_qq}
    \end{subfigure}
    \begin{subfigure}[b]{0.24\textwidth}
        \includegraphics[width=\textwidth]{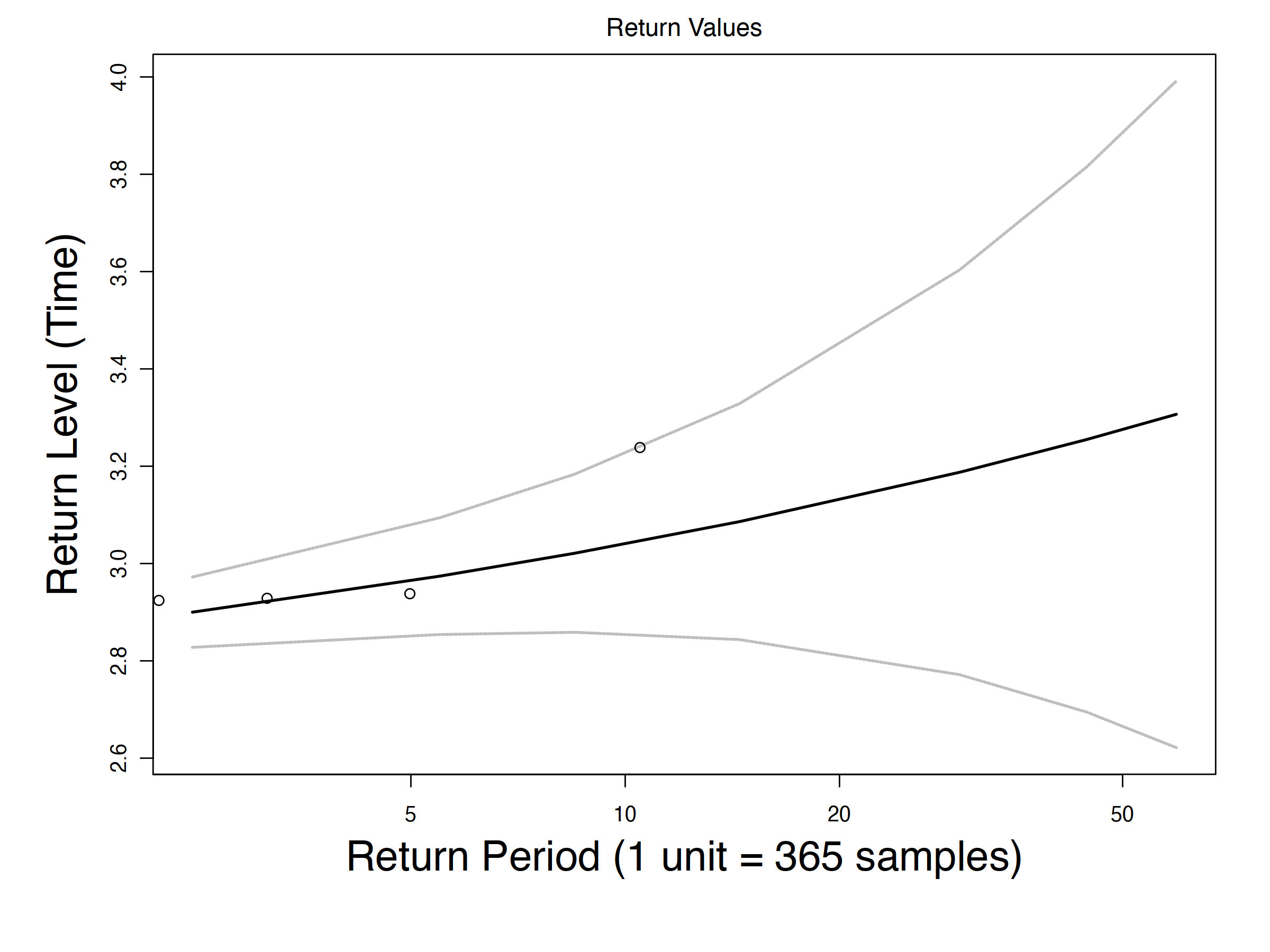}
        \caption{}
        \label{fig:DA_return_level}
    \end{subfigure}
    \caption{Linear Discriminant Analysis. (a) the computation times of training Linear Discriminant Analysis varying input dataset and hyperparameters with a threshold set to 2.8 (s),
    (b) the density plot for GEV of Linear Discriminant,
    (c) quantile Plot for the execution time of Linear Discriminant,
    (d) m-return level plot for Linear Discriminant with expected values and their 95\% CI.}
\end{figure*}

\vspace{0.25 em}
\noindent \textit{Feasibility and Scalability.} Figure~\ref{fig:DA_threshold}
shows the computation costs for LDA. We set the 
threshold for extreme training computation times to 2.8 (s).
The location, scale, and shape of GEV are 2.8, 0.1, and 0.25, respectively
(see Figure~\ref{fig:DA_density}). 
The QQ plot (see Figure~\ref{fig:DA_qq}) showed that the validity is up to 3.0 (s).

\vspace{0.25 em}
\noindent \textit{Usefulness.} For 500, 1K, and 2K return periods,
the return levels are 2.9 (s) [2.8, 3.0], 2.9 (s) [2.8, 3.0], 3.0 (s) [2.9, 3.1], and
3.1 (s) [2.8, 3.3], respectively (see Figure~\ref{fig:DA_return_level}).
The corresponding WCCTs remain at 3.2 (s) for all these periods. Thus, the actual
WCCT of 2K queries is the only one that resides within the confidence range of predicted WCCT.  
Based on the characteristics of EVT, we expect to observe an extreme convergence time of 2.9 (s) 
one in every 650 queries, and the likelihood of observing such an extreme training time is 0.2\%.

\vspace{0.5 em}
\noindent \textbf{Gaussian Process.} 
Gaussian process is a Bayesian classifier that infers posterior over linear classifier parameters using a Gaussian prior and observed data. We analyze its implementations in scikit-learn library~\cite{scikit-learn}.

\vspace{0.25 em}
\noindent \textit{Test Cases.} Similar to the previous case studies, within $4$ hours,
we train $506$ models and seek to estimate the worst-case training convergence time via EVT.

\vspace{0.25 em}
\noindent \textit{Feasibility and Scalability.} We set the threshold for extreme
training computation times to 26.6 (s). 
The location, scale, and shape of GEV are 43.0, 1.6, and -0.72, respectively. 

\vspace{0.25 em}
\noindent \textit{Usefulness.} For the 500 return period,
the return level is 43.6 (s) [42.0, 45.5]. The actual WCCT of 500 queries is
44.6 (s). We expect to observe an extreme training to converge of 44.0 (s) one in every 253 queries,
and the likelihood of observing such an extreme training time is 0.4\%. 

\begin{tcolorbox}[colback=aliceblue, boxrule=1pt,left=1pt,right=1pt,top=1pt,bottom=1pt]
\textbf{Answer RQ2:}
We found that GEV is a scalable method to infer the WCCT of ML training algorithms. 
In all 4 case studies, GEV predicts the worst-case convergence times accurately in the longest
horizon (i.e., 10K). However, in 2 cases, the GEV predictions were not accurate for a shorter period
of time (i.e., for the next 500, 1K, and 2K queries). Overall, GEV predicts the WCCT accurately
in 57\% of cases. 
\end{tcolorbox}

\subsection*{RQ3: GEV predictions of WCCT for DNN Inference.}
Our next goal is to evaluate the convergence times based on the inference of deep
neural networks. We are given a set of pre-trained deep neural networks with ReLU units
that control different cyber-physical systems (CPS). We used CPS rather than
image classifications (MNIST, CIFAR, etc.) since in CPS applications, multiple DNN-based inferences are required
to infer a decision while DNN-based image classifications are often one-step
fast process. 

In each of the benchmarks below, we have a deep neural network that takes
the state of a system as the input and infers the next control value to move toward an 
equilibrium state. We measure the convergence time of DNN to stabilize the system. 
In doing so, our goal is to model and analyze the extreme convergence times: the worst-case time taken by the
DNN controller to enter a small set containing the stable state, starting from a randomly chosen initial state.

Similar to the convergence of training algorithms, we set the threshold of extreme values
to the range from the mean of observed execution times plus one standard deviation to
the mean plus two standard deviations. We measure the accuracy by comparing the
actual convergence times to the GEV-based prediction. 

\begin{figure*}[t]
\begin{tabular}{ccc}
\includegraphics[width=0.25\textwidth]{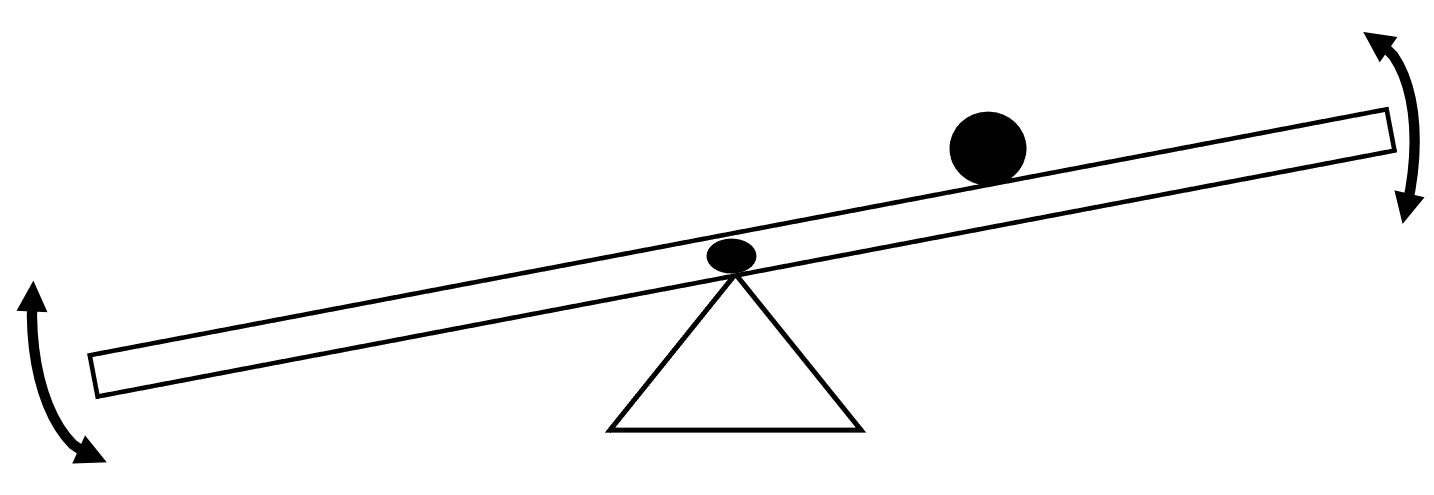} & 
\includegraphics[width=0.25\textwidth]{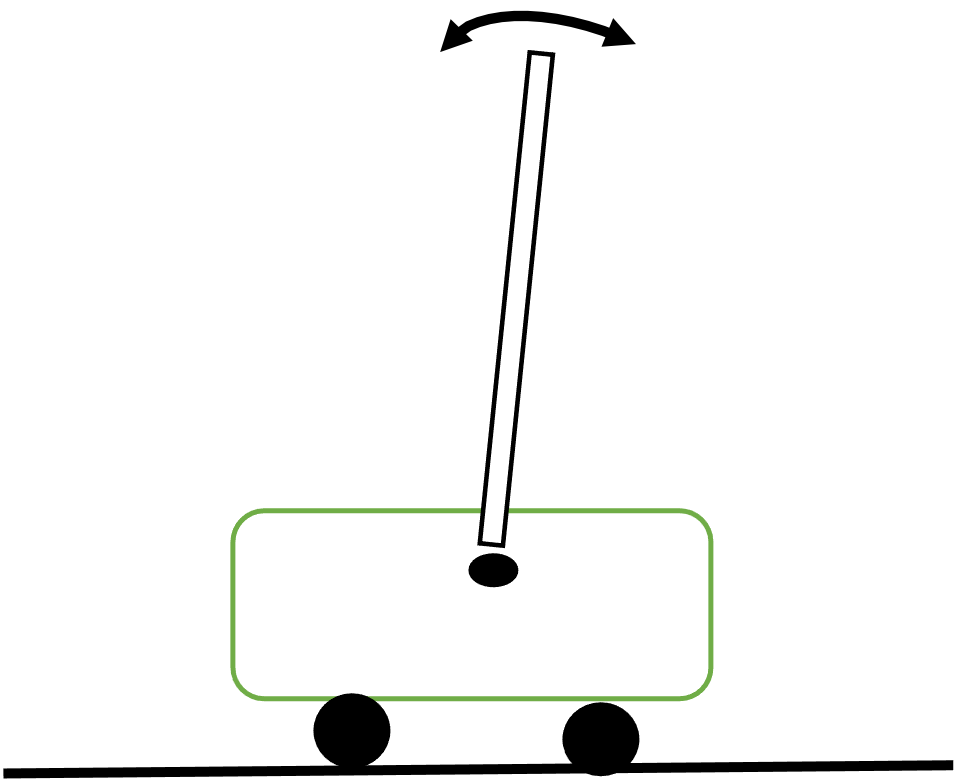} & \includegraphics[width=0.25\textwidth]{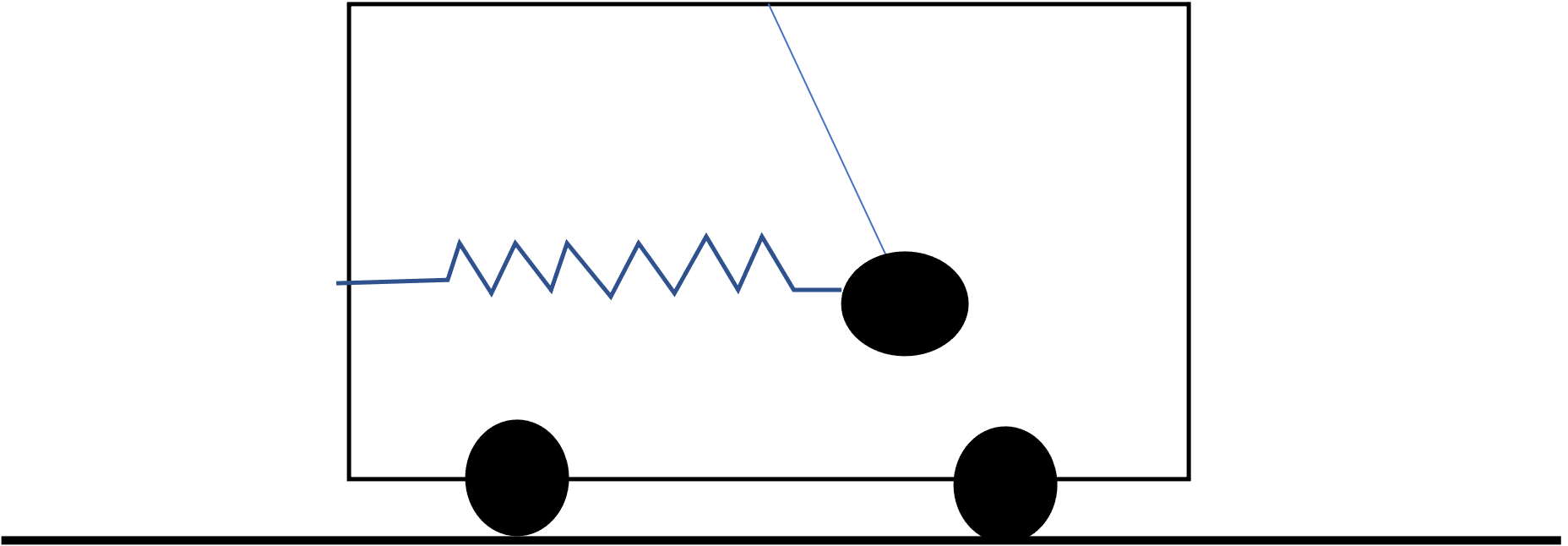}\\
(a) Ball and Beam & (b) Inverted Pendulum & (c) Tora \\  
\end{tabular}
\caption{Schematic diagrams for the neural-network controlled physical systems.}\label{fig:nn-cps-figures}
\end{figure*} 

\vspace{0.5 em}
\noindent \textbf{Ball and Beam:} Figure~\ref{fig:nn-cps-figures} (a) shows the setup for a ball-and-beam system with a beam whose tilt can be controlled by a motor and a ball which is to be brought to rest at the center of the beam. 
The CPS system has $6$ variables: $x$ denotes the displacement of the ball from the center of the beam; $v$ shows the velocity; $\theta$ is the angle of the beam; $\omega$ denotes the angular velocity of the beam; $u$ denotes the torque delivered by the motor; and $d$ denotes a random disturbance force chosen uniformly at random from the range $[-0.05,0.05]$ every $0.1$ seconds.
Given $(x, v, \theta, \omega)$ as the inputs, the control variable $u$ is inferred by a ReLU neural network with $5$ hidden layers and  $106$ neurons in total. The differential equation model for $(x, v, \theta, \omega)$, the DNN model, and the synthesis methodology of ``region stability'' are discussed in~\cite{Dutta+Others/2018/Learning}.

\vspace{0.25 em}
\noindent \textit{Test Case.} We simulate the DNN controller and generate 10,000 random test
cases. These are i.i.d. random initial states drawn from the range: $x \in [-1, 1], v \in [-1,1], \theta \in [-0.2, 0.2], \omega \in [-0.1, 0.1]$ and disturbance input as described above. Our goal is to measure how long the system takes to settle and remain inside the region $(x, v, \theta, \omega) \in [-0.05,0.05]^4$. The total time taken for 10,000 simulations is 11 mins 33 seconds (each run simulates the system for $30$ time units with a time step of $0.02$ time units).

\begin{figure*}[tbp!]
    \centering
    \begin{subfigure}[b]{0.24\textwidth}
        \includegraphics[width=\textwidth]{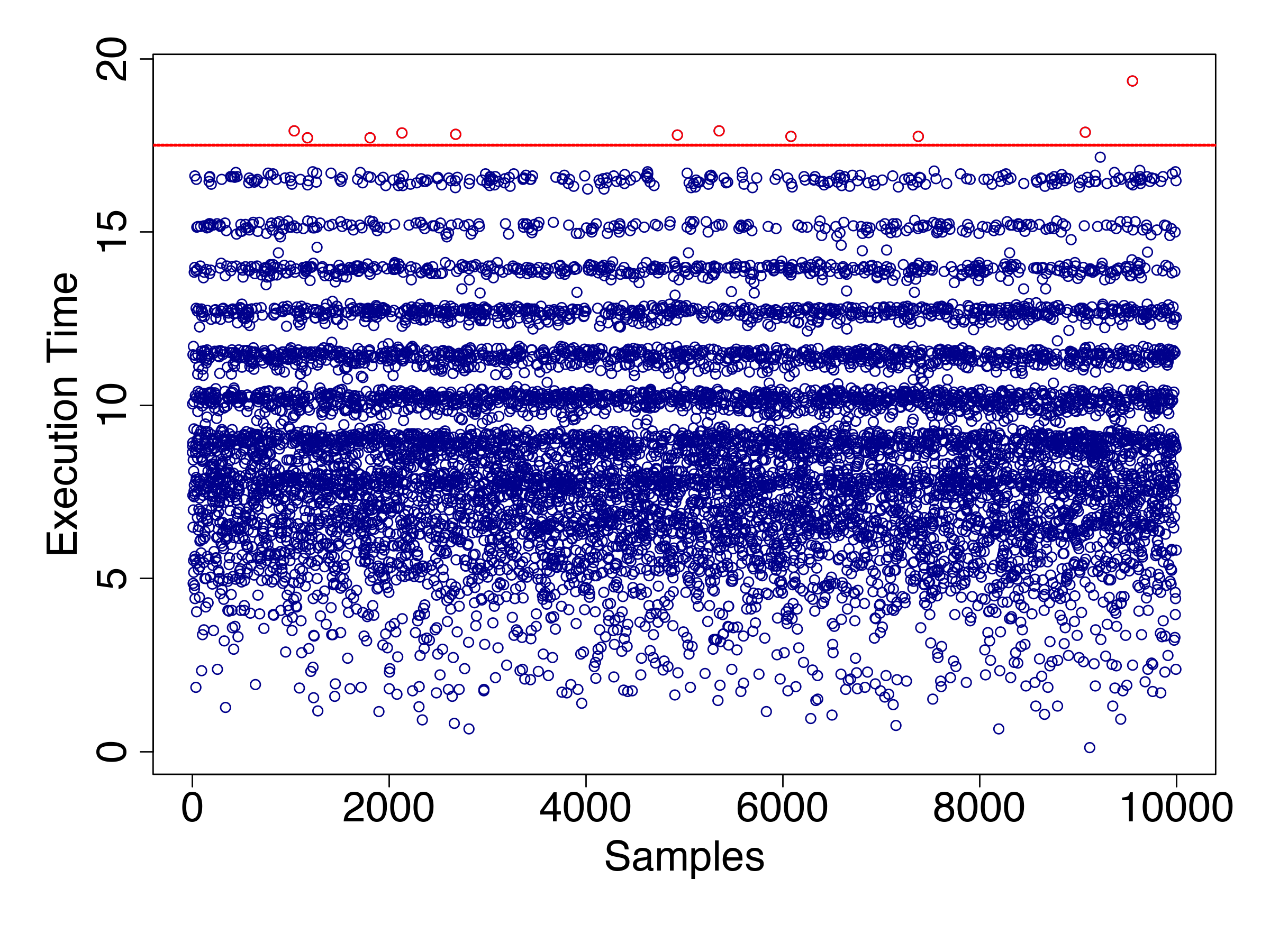}
        \caption{}
        \label{fig:BB_threshold}
    \end{subfigure}
    \begin{subfigure}[b]{0.24\textwidth}
        \includegraphics[width=\textwidth]{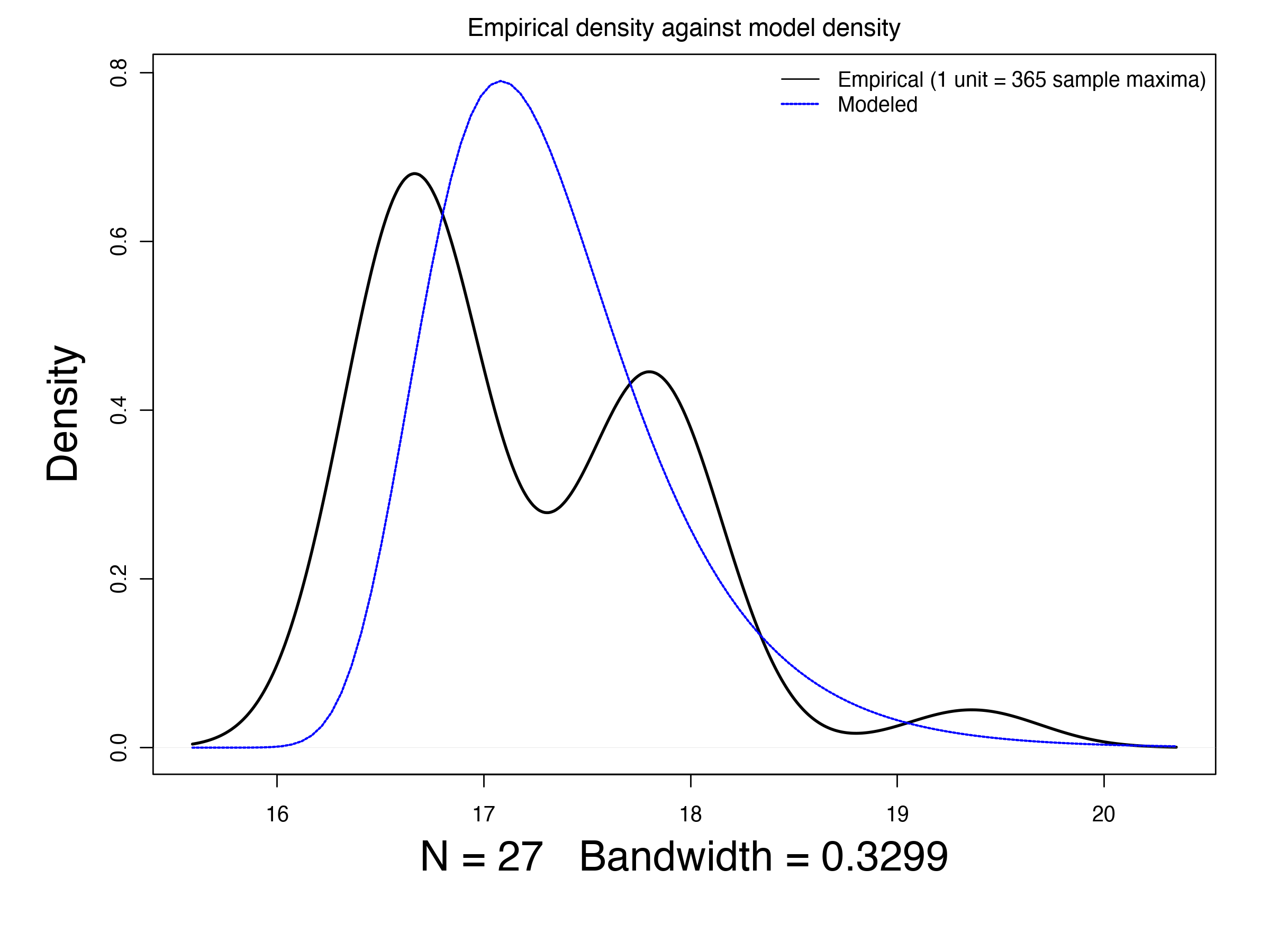}
        \caption{}
        \label{fig:BB_density}
    \end{subfigure}
    \begin{subfigure}[b]{0.24\textwidth}
        \includegraphics[width=\textwidth]{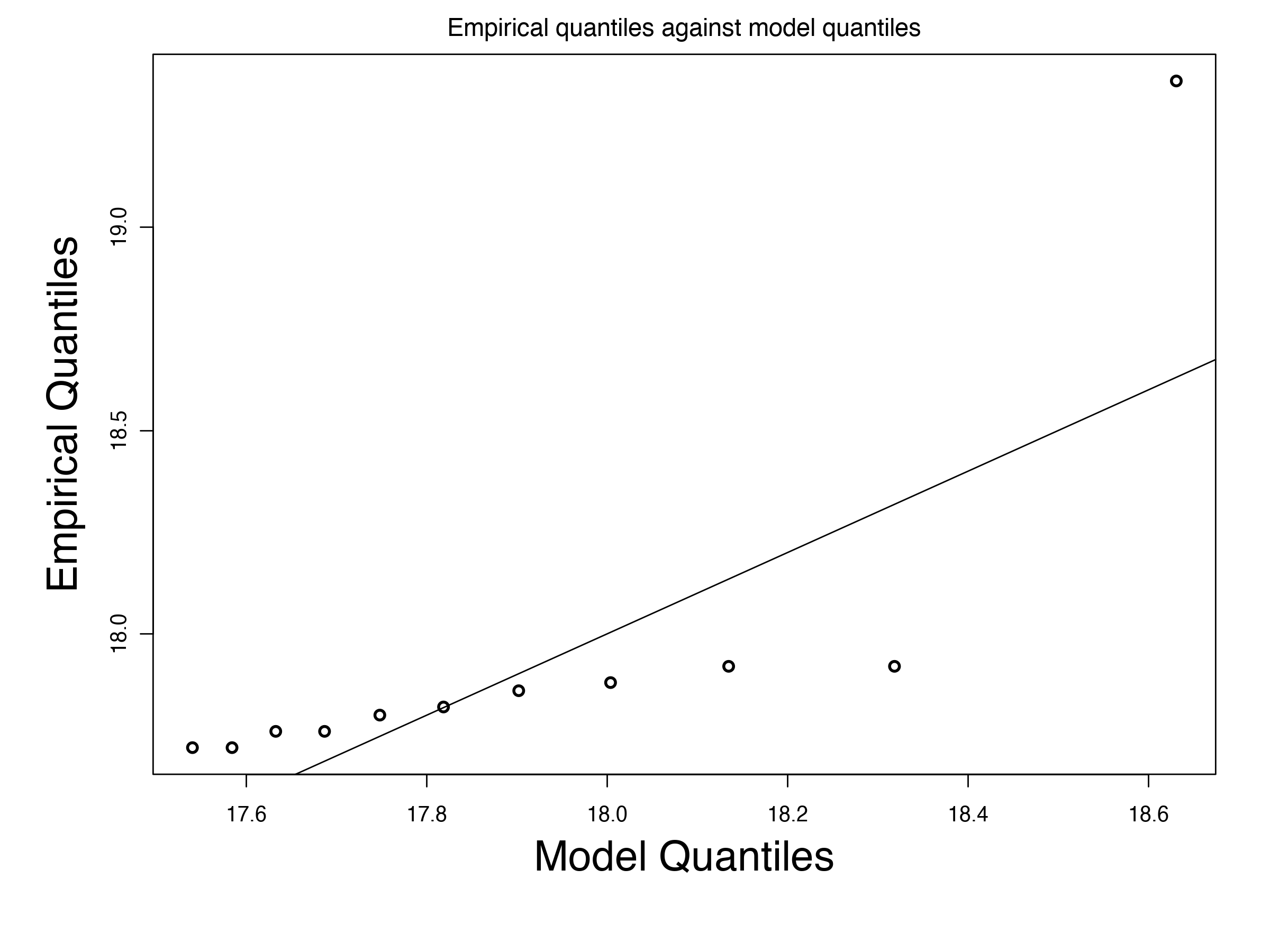}
        \caption{}
        \label{fig:BB_qq}
    \end{subfigure}
    \begin{subfigure}[b]{0.24\textwidth}
        \includegraphics[width=\textwidth]{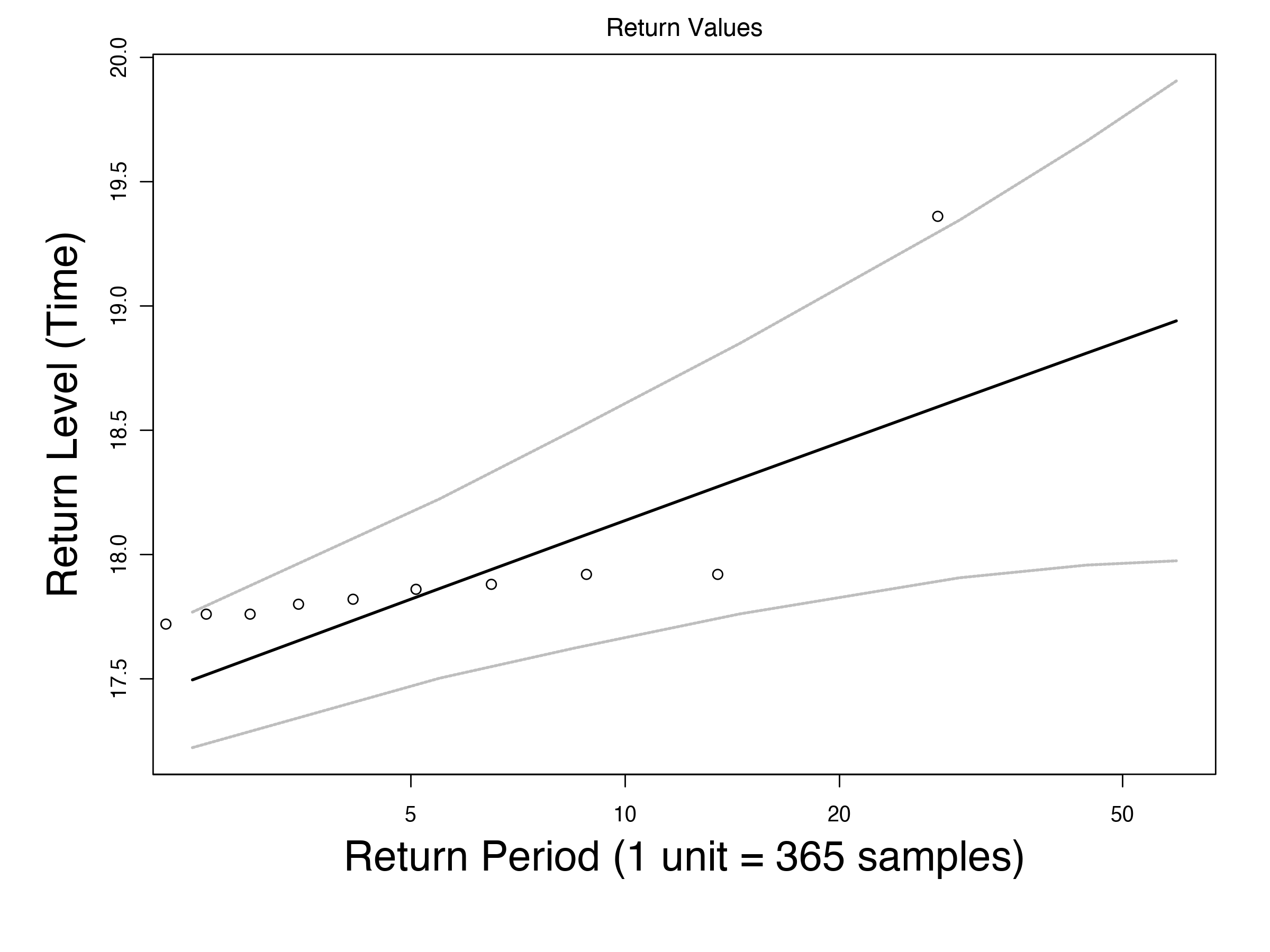}
        \caption{}
        \label{fig:BB_return_level}
    \end{subfigure}
\caption{Ball-beam. (a) the converge times of Ball-Beam on the randomly generated inputs,
(b) quantile Plot for the execution time of Ball-Beam,
(c) the density plot for GEV of Ball-Beam,
(d) m-return level plot for Ball-Beam with expected values and their 95\% CI.}
\end{figure*}

\vspace{0.25 em}
\noindent \textit{Feasibility and Scalability.}
Figure~\ref{fig:BB_threshold} shows the recorded settling times as well as the threshold of $17.5$ for extreme values.
It takes $2$ seconds to infer GEV distributions. Figure~\ref{fig:BB_density} shows
the empirical vs. modeled probability density functions. The location, scale, and
shape of GEV are 17.1, 0.5, and 0.0, respectively. The shape shows that GEV is
type I, and the density of tail distribution is infinite, but decaying exponentially.
Figure~\ref{fig:BB_qq} shows the QQ plot which is used to find the bounds on the 
extrapolations. 

\vspace{0.25 em}
\noindent \textit{Usefulness.}
Figure~\ref{fig:BB_return_level} shows the return levels
and their 95\% confidence intervals. For 1K, 2K, 5K, and 10K return periods,
the return levels are 17.4 (s) [17.1, 17.7], 17.8 (s) [17.5, 18.2], 
18.3 (s) [17.7, 18.8], and 18.6 [17.9, 19.3], respectively. The actual WCCTs 
are 17.8, 17.8, 19.3, and 19.3, for 1K, 2K, 5K, and 10K simulations. Those,
predictions for 2K and 10K simulations are predicted accurately. 
The expected likelihood of observing the settling
times above 17.4 (s) and 17.8 (s) after 100th queries are 10.4\% and 6.0\%, respectively.

%  Cross-out for now: bring it back

\vspace{0.5 em}
\noindent \textbf{Inverted Pendulum} We consider the control of an inverted pendulum on a cart (see Figure~\ref{fig:nn-cps-figures} (b)) using a neural networked controlled system.
The key idea is to stabilize an inverted pendulum by moving the cart back and forth so that the entire system comes to rest with the pendulum pointing directly upwards. This is a very commonly studied control problem with numerous applications 
such as segway scooters.
We consider a reduced nonlinear model with two state-variables: $\theta$ denoting the angle of the pendulum wherein $0$ radians denote an upright pendulum and $\omega$ denotes the angular velocity. 
\[ \dot{\theta} = \omega,\ \dot{\omega} = \sin(\theta) - (u + d) \cos(\theta) \,.\] 
Here $u$ is the control input and $d$ is a random disturbance chosen every $0.1$ seconds uniformly from the range $[-0.1,0.1]$. Starting from initial conditions $(\theta, \omega) \in [-1,1]^2$, we are interested in the time taken by a neural network controller to settle and remain in the range $[-0.05, 0.05]^2$. The neural network controller has two hidden layers with a total of $20$ neurons.

\vspace{0.25 em}
\noindent \textit{Test Case.} We interact with DNN controller and generate 10,000 random test
cases with initial states and disturbance inputs drawn uniformly from the ranges described above. The total time taken for 10,000 simulations is 2 minutes and 35 seconds (each run simulates the system for $30$ time units with a time step of $0.02$ time units).

\begin{figure*}[tbp!]
    \centering
    \begin{subfigure}[b]{0.24\textwidth}
        \includegraphics[width=\textwidth]{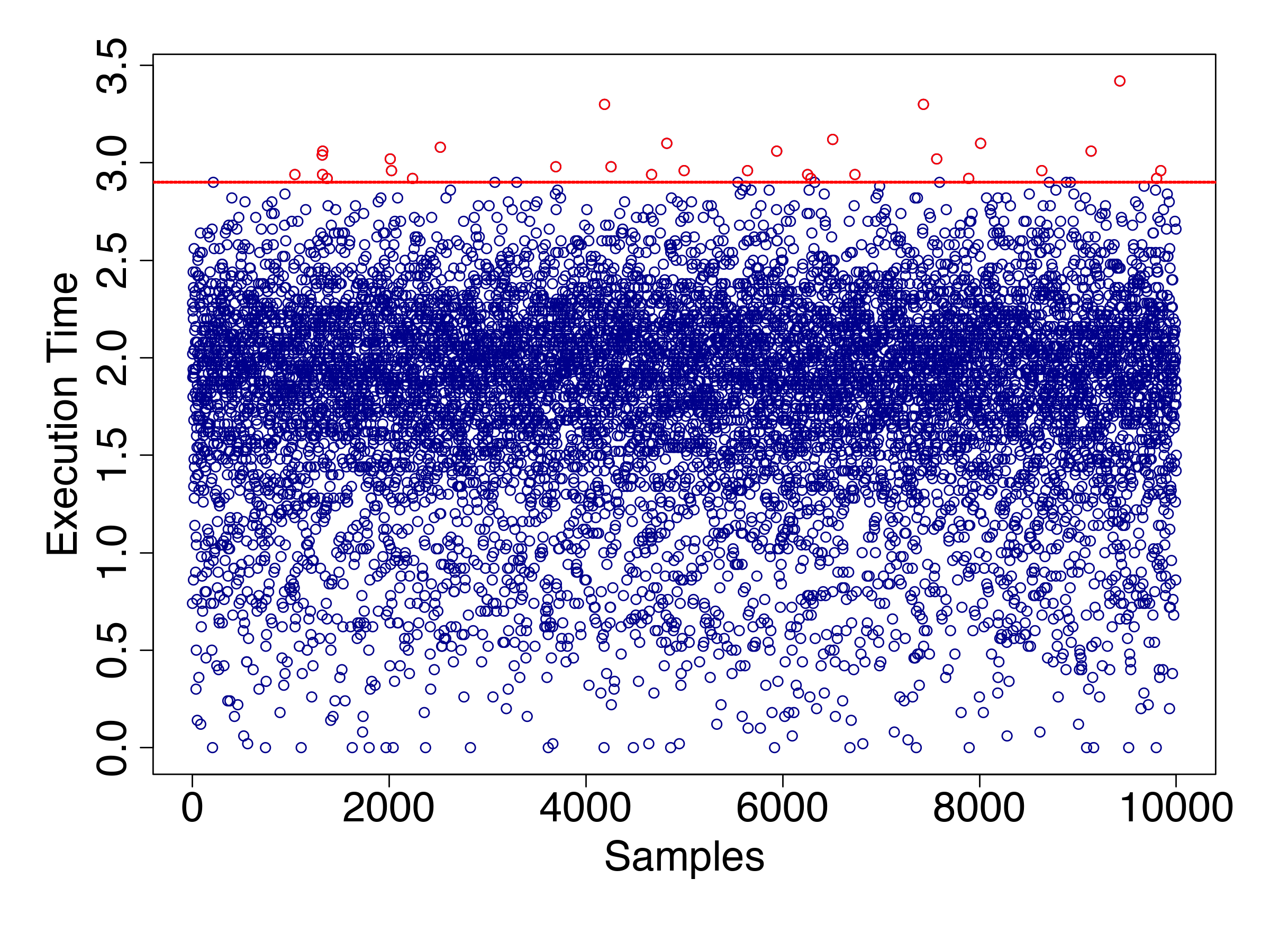}
        \caption{}
        \label{fig:SP_threshold}
    \end{subfigure}
    \begin{subfigure}[b]{0.24\textwidth}
        \includegraphics[width=\textwidth]{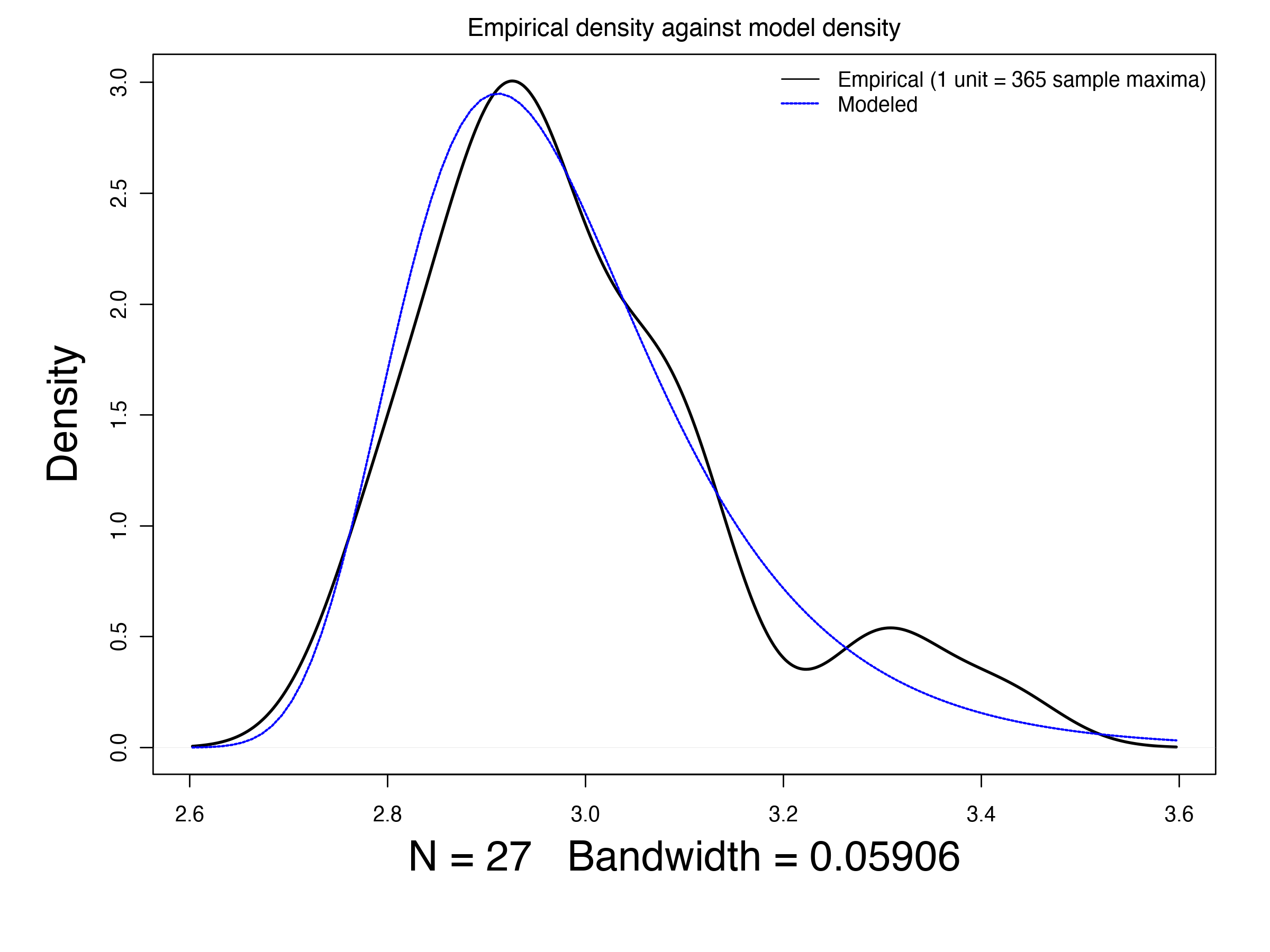}
        \caption{}
        \label{fig:SP_density}
    \end{subfigure}
    \begin{subfigure}[b]{0.24\textwidth}
        \includegraphics[width=\textwidth]{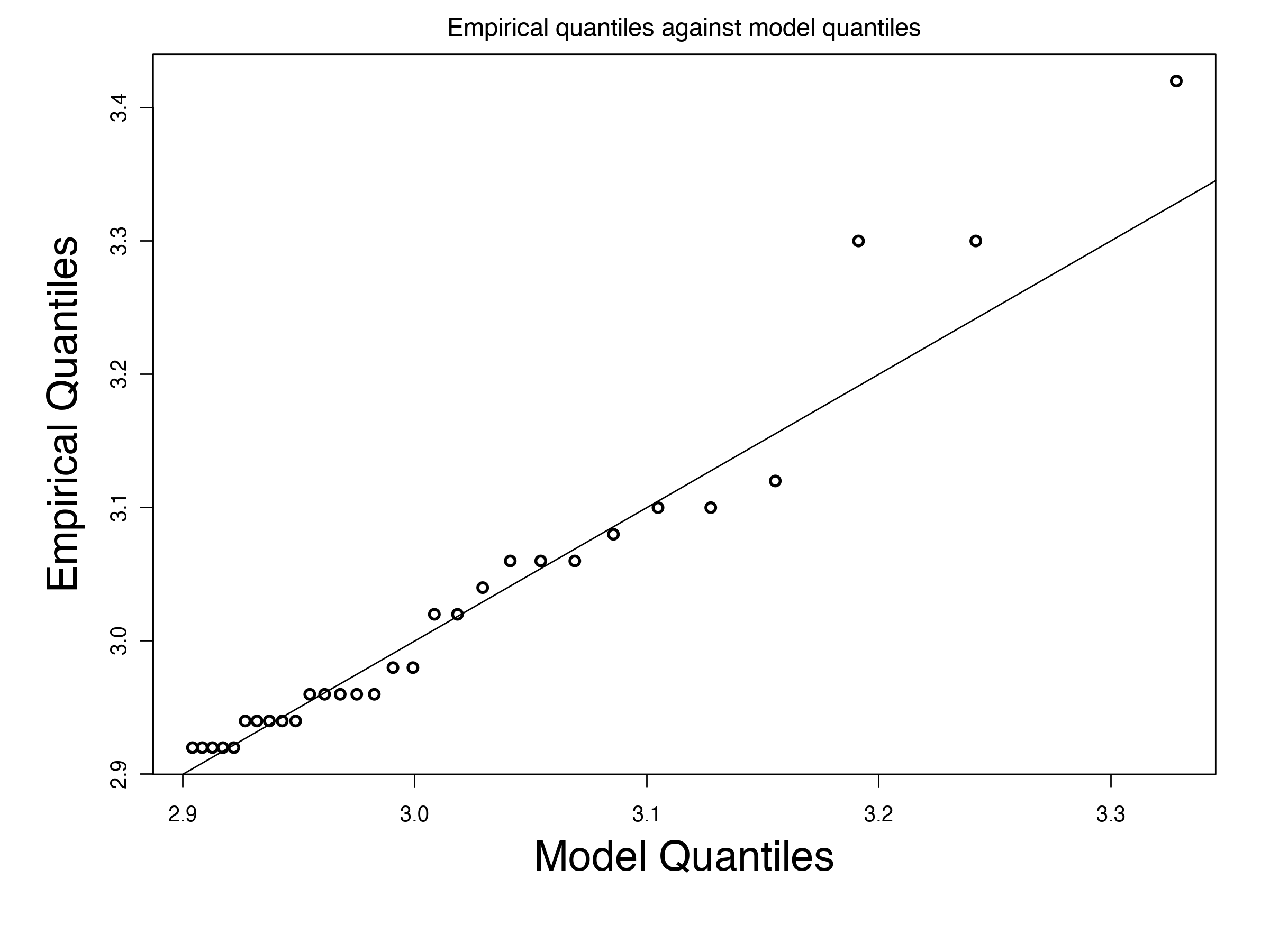}
        \caption{}
        \label{fig:SP_qq}
    \end{subfigure}
    \begin{subfigure}[b]{0.24\textwidth}
        \includegraphics[width=\textwidth]{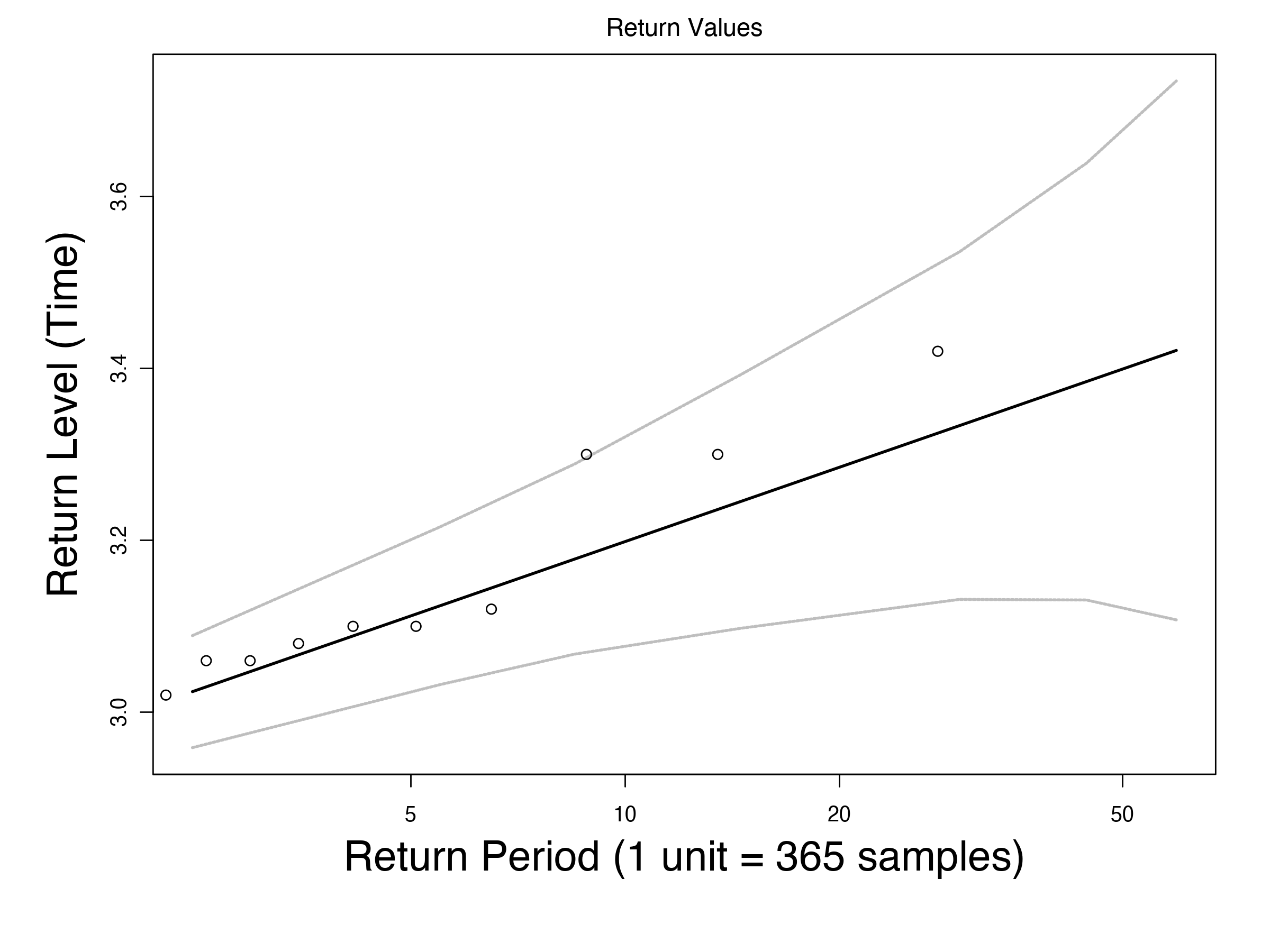}
        \caption{}
        \label{fig:SP_return_level}
    \end{subfigure}
\caption{Inverted Pendulum. (a) the converge times of Pendulum on the randomly generated inputs,
(b) quantile Plot for the execution time of Pendulum,
(c) the density plot for GEV of Pendulum,
(d) m-return level plot for Pendulum with expected values and their 95\% CI.}
\end{figure*}

\vspace{0.25 em}
\noindent \textit{Feasibility and Scalability.}
Figure~\ref{fig:SP_threshold}
shows the recorded settling times as well as the threshold of $2.9$ for extreme values.
It takes $2$ seconds to infer GEV distributions. Figure~\ref{fig:SP_density} shows
the empirical vs. modeled probability density functions. The location, scale, and
shape of GEV are 2.9, 0.1, and 0.0, respectively. The shape shows that GEV is
type I, and the density of tail distribution is infinite, but decaying exponentially.
Figure~\ref{fig:SP_qq} shows the QQ plot which might be valid up to 3.3 (s).

\vspace{0.25 em}
\noindent \textit{Usefulness.}
Figure~\ref{fig:SP_return_level} shows the return level
plots and its 95\% confidence intervals. For 1K, 2K, 5K, and 10K return periods,
the return levels are  3.0 (s) [2.9, 3.1], 3.1 (s) [3.0, 3.2], 3.2 (s) [3.1, 3.4],
and 3.3 (s) [3.1 3.5], respectively. The actual WCCTs are 3.0 (s), 3.0 (s), 3.3 (s),
and 3.4 (s) for 1K, 2K, 5K, and 10K simulations, respectively. The predictions include
the actual WCCT for all the simulation queries. 
The expected likelihood of observing the computation
times above 3.0 (s) after 100th queries is 12.2\%.

%  Cross-out for now: bring it back

\vspace{0.5 em}
\noindent \textbf{Tora:} Figure~\ref{fig:nn-cps-figures} (c) shows the TORA (Translational Oscillator with Rotational Actuator), a widely studied nonlinear control model~\cite{Robert+Others/1998/Benchmark}. Its simplified model involves $4$ state variables $(x_1, \ldots, x_4)$ whose dynamics are given by:
\[  \dot{x_0} = x_1,\ \dot{x_1} = -x_0 + 0.1 \sin (x_2) +d,\ 
    \dot{x_2} = x_3,\ \dot{x_3} = u\,.\]
The initial states are drawn from ranges $(x_1, x_2) \in [-1, 1]^2 $ and $(x_3, x_4) \in [-0.5, 0.5]^2$ with the disturbance inputs $d$ drawn from the range $[-0.01,0.01]$. The settling region is taken to be $[-0.1, 0.1]^4$.
The neural network has a single hidden layer with just one neuron.

\vspace{0.25 em}
\noindent \textit{Test Case.} We perform $10,000$ random simulations of the system, requiring a total of 40 minutes and 50 seconds (each run simulates the system for $300$ time units with a time step of $0.02$ time units).

\begin{figure*}[tbp!]
    \centering
    \begin{subfigure}[b]{0.24\textwidth}
        \includegraphics[width=\textwidth]{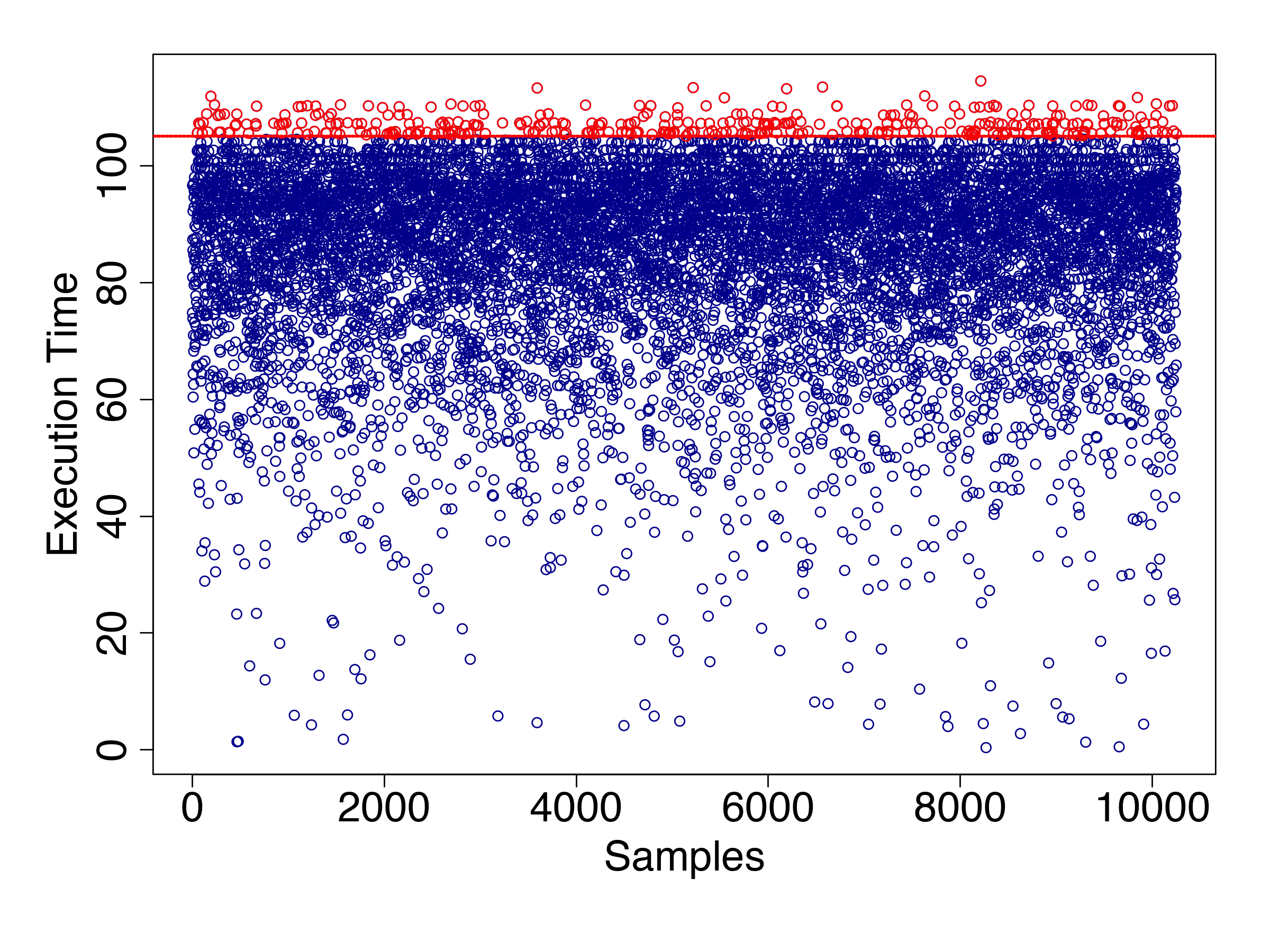}
        \caption{}
        \label{fig:To_threshold}
    \end{subfigure}
    \begin{subfigure}[b]{0.24\textwidth}
        \includegraphics[width=\textwidth]{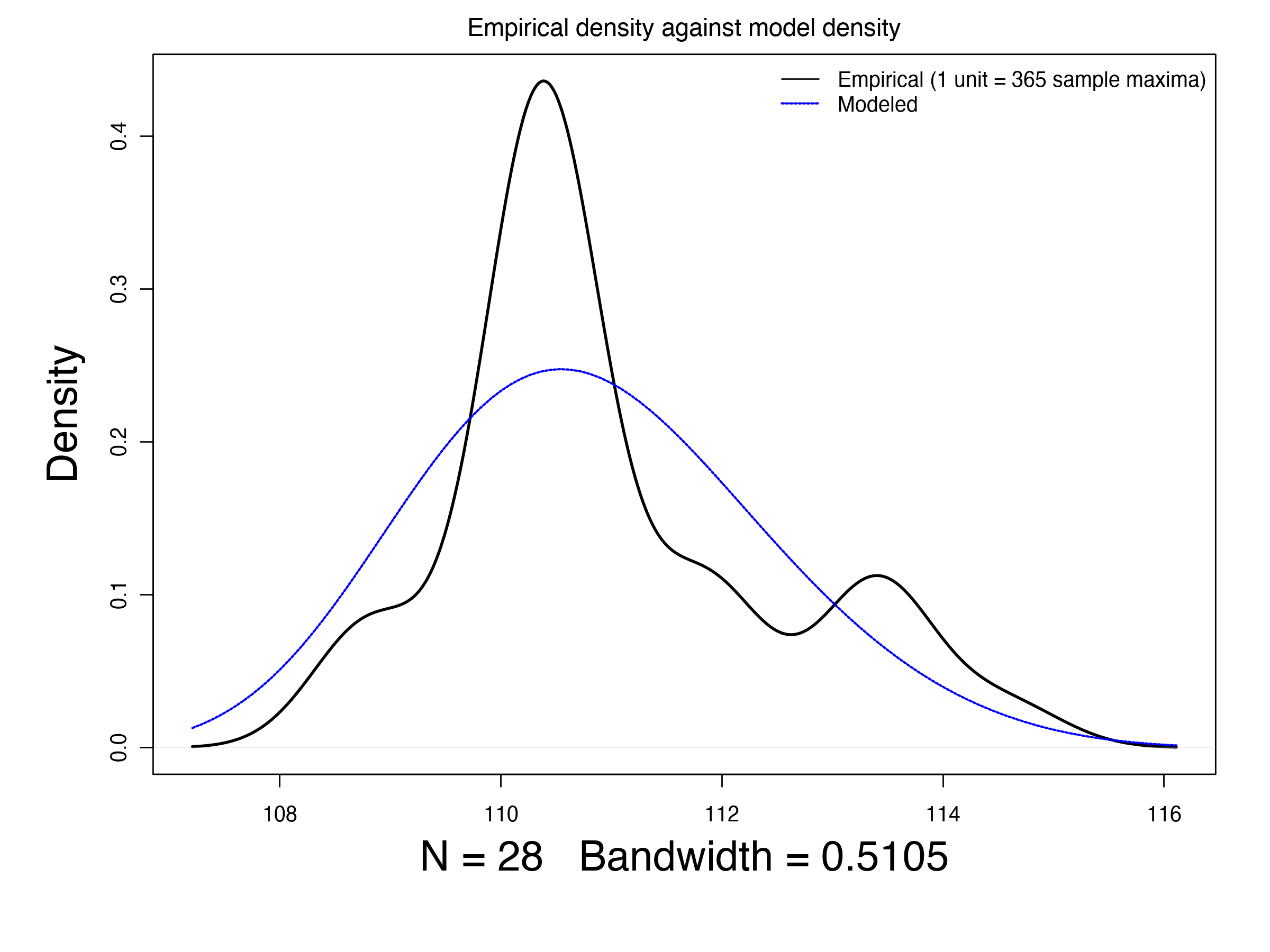}
        \caption{}
        \label{fig:To_density}
    \end{subfigure}
    \begin{subfigure}[b]{0.24\textwidth}
        \includegraphics[width=\textwidth]{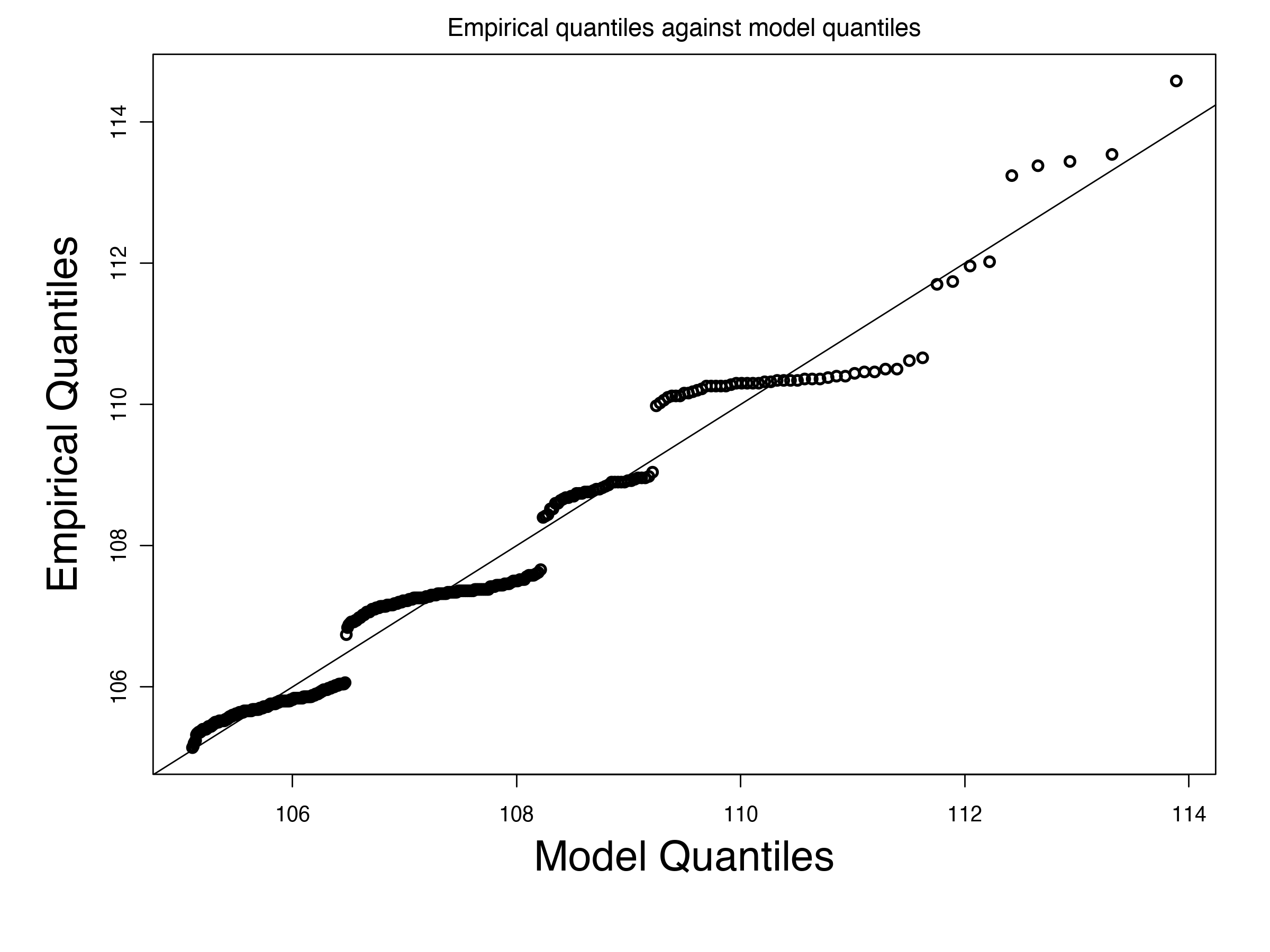}
        \caption{}
        \label{fig:To_qq}
    \end{subfigure}
    \begin{subfigure}[b]{0.24\textwidth}
        \includegraphics[width=\textwidth]{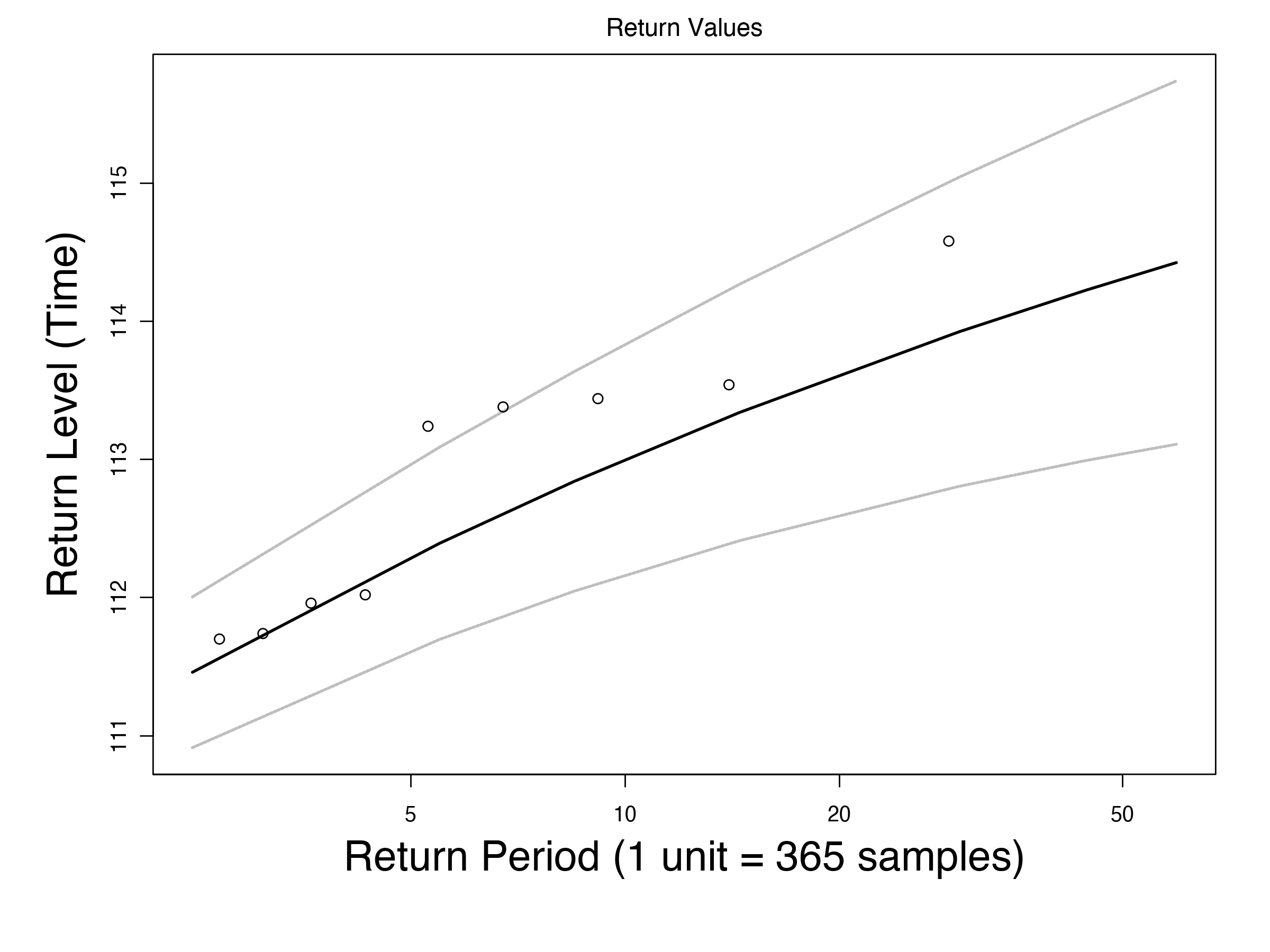}
        \caption{}
        \label{fig:To_return_level}
    \end{subfigure}
\caption{Tora. (a) the converge times of Tora on the randomly generated inputs,
(b) quantile Plot for the execution time of Tora,
(c) the density plot for GEV of Tora,
(d) m-return level plot for Tora with expected values and their 95\% CI.}
\end{figure*}

\vspace{0.25 em}
\noindent \textit{Feasibility and Scalability.}
Figure~\ref{fig:To_threshold}
shows the recorded computation times as well as the threshold of $105.1$ for extreme values.
It takes $2$ seconds to infer GEV distributions. Figure~\ref{fig:To_density} shows
the empirical vs. modeled probability density functions. The location, scale, and
shape of GEV are 110.2, 1.5, and 0.0, respectively. The shape shows that GEV is
type I and the density of tail distribution is finite.
Figure~\ref{fig:To_qq} shows the QQ plot.

\vspace{0.25 em}
\noindent \textit{Usefulness.}
Figure~\ref{fig:To_return_level} shows the return levels
and their 95\% confidence intervals. For 1K, 2K, 5K, and 10K return periods,
the return level is 111.6 (s) [110.8, 113.3], 112.4 (s) [111.5, 113.3], 
113.3 (s) [112.1, 114.5], and 113.9 [112.5, 115.3], respectively.
The actual WCCTs have remained 113.5 (s). Therefore, the predictions
up to 5K and 10K simulations are accurate. The expected likelihood to observe the computation
times above 111.3 (s) after 100th interaction is 8.4\%.

%  Cross-out for now: bring it back

\begin{tcolorbox}[colback=aliceblue, boxrule=1pt,left=1pt,right=1pt,top=1pt,bottom=1pt]
\textbf{Answer RQ3:}
We found that GEV is a feasible and scalable method to infer the WCCT of DNN inferences for cyber-physical
systems. In all 3 case studies, GEV predicts the worst-case convergence times accurately in the longest
horizon (i.e., 5K and 10K), similar to the case studies for the ML training algorithm. Consider all 12
predictions, we observe that GEV predictions are accurate for 9 cases, with an overall accuracy of 75\%. 
\end{tcolorbox}

\section{Discussions}
\label{sec:discussion}

EVT only considers the external manifestations of a system and does not provide information on how the input observations are made. 
To ensure the theory applies, it is necessary to understand the execution conditions that programs may experience at deployment time and control the execution conditions during analysis time to ensure they are representative. 
In addition, EVT does not guarantee the representativeness of the data, which depends on the quality of test cases and
the state of the environment. The WCCT estimates obtained with EVT are only valid for the data population sampled or the observed operating conditions. If representativeness is low, the WCCT bounds may not be a reliable prediction. 

\noindent \textit{Internal Validity.}
The EVT applies when the observations are independent and identically distributed. While leveraging
DPFuzz, we mitigate the dependency between inputs by randomizing observations and controlling the size of inputs. 
In addition, the threshold of EVT should be chosen judiciously, otherwise the
GEV distributions might include non-tail samples (mixture distributions)
or a few tail samples that cast doubts on the confidence. While we used
the characteristics and confidence of EVT to choose a threshold; more
research is needed to find principles for picking the threshold values
for the WCCT analysis of ML-based software to achieve a higher accuracy/precision
in the analysis. Moreover, the number of samples to fit into the EVT engine is
also important for a precise analysis. While we used the Bayes factor
to decide on the number of samples for EVT analysis,
it requires further research. 

\noindent \textit{External Validity.}
To ensure that the results are generalizable, we consider multiple case studies for both training
and inference of ML algorithms. 
While we consider deep neural networks, we only consider the pre-trained DNNs as CPS controllers. We left
studying the worst-case convergence of training DNNs for future work.

\section{Related Work}
\label{sec:related}
\noindent \textbf{Extreme Value Theory for Probabilistic WCET.} Extreme value theory has been significantly adapted to provide probabilistic guarantees on the worst-case execution times in the real-time and embedded systems~\cite{lu2011new,cucu2012measurement,hansen2009statistical}. 
Lima et al.~\cite{7557881} studied the applicability of extreme value theory for estimating worst-case execution times in an embedded platform equipped with a random cache of configurable sizes. Similar to us, they found that the worst-case execution times can be modeled as one of the three extreme value distributions (Weibull, Gumbel or Frechet), and thus restricting the analysis to one distribution can lead to poor or even unsafe WCET estimates. They extensively studied the effects of cache over the timing observations of embedded systems, while we focused on varying the inputs to the training process and inference models of machine learning software. 
Santinelli et al.~\cite{santinelli2014sustainability} studied the impacts of relaxing i.i.d. assumptions in the EVT modeling of WCET. They offered various tests to understand whether samples are independent and identical. They show the limitations of EVT in such cases and propose techniques such as bootstrapping to convey independence. 
Lima and Bate~\cite{lima2017valid} tackle the inherent problems of time measurements over modern hardware for the EVT analysis. They proposed indirect estimation in statistical time analysis (IESTA), which is based on randomizing the data to untangle the dependency on the hardware.
The effectiveness of IESTA is demonstrated through experiments on two real case studies involving execution time measurements from an embedded platform and a Rolls-Royce Full Authority Digital Engine Controller.
Cazorla et al.~\cite{cazorla2013upper} presents a framework to discuss the limitation of EVT in computing upper bounds on the execution time of programs. Specifically, they consider the problem of sufficiency of observational data to infer extreme value distributions. Their recommendations include controlling the execution conditions at analysis time and understanding the representativeness of the analysis-time execution conditions with respect to those that may occur during operation. They identify various sources of variability of execution times and propose a framework to sample inputs randomly while ensuring representative maxima samples from those sources of variability. Since identifying all of the sources of variability in our applications is hard, we are required to develop new techniques to understand the representativeness of data. 
Overall, all previous work focused on real-time and embedded systems, while we measured the convergence time of ML algorithms.

\vspace{0.25 em}
\noindent \textbf{Extreme Value Theory for Detecting Rare Bugs in Circuit Design.}
Statistical blockage~\cite{singhee2007statistical} used EVT to block unwanted rare
events to improve circuit reliability. In particular, the authors found standard MCMC techniques
such as importance sampling are inefficient in modeling unlikely rare events. Instead,
they used MCMC sampling to infer the parameters of EVT distributions.
Antoniadis et al.~\cite{8342220} adapted EVT to estimate the worst-case delay of VLSI circuits under
variations in gate/interconnect parameters. 
Cooley et al.~\cite{doi:10.1198/016214506000000780} used the generalized Pareto distribution
to predict flooding based on daily precipitations above a high threshold. A similar
analysis has proposed to study how a slowly changing climate would possibly lead to more
frequent extreme events~\cite{cooley2009extreme}. EVT tools have been also used
to answer hypothetical questions about physical endurance~\cite{EVT-data-article}.
We refer interested
readers to classic EVT books to see other examples and applications~\cite{coles2001introduction,de2006extreme,Davison1984}.

\vspace{0.25 em}
\noindent \textbf{Characterization of WCET through program analysis.} 
At a high-level, there are two major approaches to estimating the worst-case
execution times in software: static analysis and dynamic analysis.
Here, we discuss these techniques as well as those that combine these
two techniques.

\noindent \textit{Static Analysis.} \textsc{SAFER}~\cite{2009-Inputs} combines
taint analysis with control dependency analysis to identify high-complexity control structures
whose execution can lead to resource exhaustion such as CPU clocks and stack space.
While these techniques can detect infinite executions, they rely on expensive
taint analysis that might not be feasible for some real-world applications (e.g., Java programs) with
dynamic features such as reflections~\cite{landman2017challenges}. In addition,
\textsc{SAFER} only detects DoS vulnerabilities, not super-linear computational
complexities.

\noindent \textit{Dynamic Analysis.} Search-based software testing has been significantly used to model the execution times~\cite{petsios2017slowfuzz,lemieux2018perffuzz,tizpaz2020detecting,tizpaz2017discriminating,Tizpaz-Niari_Cerny_Chang_Trivedi_2018,noller2021qfuzz}.
For example, \textsc{PerfFuzz}~\cite{lemieux2018perffuzz} adapts evolutionary algorithms to maximize the
cost of different entities in the control-flow graph (such as the number of times to take an edge) that lead
to precise characterization of WCET in large-scale systems. These techniques often
discover a single input that characterizes WCET whereas EVT techniques provide
rich statistical information such as the expected worst-case execution times,
the return levels of WCET, and their likelihood. We adapted \textsc{DPFuzz}~\cite{10.1145/3406882},
a similar ML-oriented fuzz testing, to generate test cases.

\noindent \textit{Hybrid Analysis.} \textsc{GameTime}~\cite{seshia2011gametime} combines
basis path analysis through SMT solvers with random suite test generations to predict
various timing properties of software including WCET. In particular, they use CFG
paths with total execution times to infer the computation times of each element in the CFG.

\section{Conclusion and Future Work}
\label{sec:conclusion}
Timing analysis is a crucial non-functional property of ML-based software systems
but poses significant challenges to traditional static and dynamic program analysis
methods. We proposed a tool and a technique based on the statistics of extreme value theory to model 
the worst-case convergence time of ML algorithms.
Our experiments showed that EVT-based WCCT analysis is feasible,
scalable, and accurate for the timing analysis of ML algorithms.

Our observations include that i) EVT becomes more accurate in the longer
horizon than the shorter period of time and ii) EVT was more accurate in
predicting the DNN inference than predicting the ML training convergence
times.  There are also multiple other exciting future directions. 
One direction is to infer classifiers from the characteristics
of inputs (and their features) that might manifest the worst-case convergence times
as a filtering mechanism, especially for ML-as-service frameworks.
Another direction is to incorporate EVT to validate the efficacy of
repair (mitigation) applied to fix a performance bug or improve efficiency. 

\vspace{0.5 em}
\noindent \textbf{Acknowledgement.}
The authors thank the anonymous CAIN reviewers for their comments to improve this paper. 
Tizpaz-Niari partially supported by the NSF under grants CNS-2230060.

\bibliographystyle{splncs04}
\bibliography{papers}

\end{document}